\documentclass[prl,twocolumn,amssymb,showpacs]{revtex4-1}

\usepackage{graphicx}
\usepackage{color}
\usepackage{epsfig}
\usepackage{latexsym}
\usepackage{bm}

\begin{document}

\renewcommand{\ni}{{\noindent}}
\newcommand{\dprime}{{\prime\prime}}
\newcommand{\be}{\begin{equation}}
\newcommand{\ee}{\end{equation}}
\newcommand{\bea}{\begin{eqnarray}}
\newcommand{\eea}{\end{eqnarray}}
\newcommand{\nn}{\nonumber}
\newcommand{\bk}{{\bf k}}
\newcommand{\bQ}{{\bf Q}}
\newcommand{\q}{{\bf q}}
\newcommand{\s}{{\bf s}}
\newcommand{\bN}{{\bf \nabla}}
\newcommand{\bA}{{\bf A}}
\newcommand{\bE}{{\bf E}}
\newcommand{\bj}{{\bf j}}
\newcommand{\bJ}{{\bf J}}
\newcommand{\bs}{{\bf v}_s}
\newcommand{\bn}{{\bf v}_n}
\newcommand{\bv}{{\bf v}}
\newcommand{\la}{\langle}
\newcommand{\ra}{\rangle}
\newcommand{\dg}{\dagger}
\newcommand{\br}{{\bf{r}}}
\newcommand{\brp}{{\bf{r}^\prime}}
\newcommand{\bq}{{\bf{q}}}
\newcommand{\hx}{\hat{\bf x}}
\newcommand{\hy}{\hat{\bf y}}
\newcommand{\bS}{{\bf S}}
\newcommand{\cU}{{\cal U}}
\newcommand{\cD}{{\cal D}}
\newcommand{\bR}{{\bf R}}
\newcommand{\pll}{\parallel}
\newcommand{\sumr}{\sum_{\vr}}
\newcommand{\cP}{{\cal P}}
\newcommand{\cQ}{{\cal Q}}
\newcommand{\cS}{{\cal S}}
\newcommand{\ua}{\uparrow}
\newcommand{\da}{\downarrow}
\newcommand{\red}{\textcolor {red}}

\def\lsim {\protect \raisebox{-0.75ex}[-1.5ex]{$\;\stackrel{<}{\sim}\;$}}
\def\gsim {\protect \raisebox{-0.75ex}[-1.5ex]{$\;\stackrel{>}{\sim}\;$}}
\def\lsimeq {\protect \raisebox{-0.75ex}[-1.5ex]{$\;\stackrel{<}{\simeq}\;$}}
\def\gsimeq {\protect \raisebox{-0.75ex}[-1.5ex]{$\;\stackrel{>}{\simeq}\;$}}


\title{ Approximate thermodynamic structure for driven lattice gases in contact }

\author{Punyabrata Pradhan, Robert Ramsperger, and Udo Seifert}

\affiliation{ II. Institut f\"ur Theoretische Physik, Universit\"at Stuttgart,
Stuttgart 70550, Germany }

\begin{abstract}

\noindent{ For a class of nonequilibrium systems, called driven lattice gases, we study what happens when two systems are kept 
in contact and allowed to exchange particles with the total number of particles conserved. Both for attractive and repulsive 
nearest-neighbor interactions among particles and for a wide range of parameter values, we find that, to a good approximation, 
one could define an intensive thermodynamic variable, like equilibrium chemical potential, which determines the final steady 
state for two initially separated driven lattice gases brought into contact. However, due to nontrivial contact dynamics, there 
are also observable deviations from this simple thermodynamic law. To illustrate the role of the contact dynamics, we study a 
variant of the zero range process and discuss how the deviations could be explained by a modified large deviation principle.
We identify an additional contribution to the large deviation function, which we call the excess chemical potential, for
the variant of the zero range process as well as the driven lattice gases. The
excess chemical potential depends on the specifics of the contact dynamics and is in general a-priori unknown. 
A contact dependence implies that even though an intensive variable may equalize, the zeroth law could still be violated. 
}

\typeout{polish abstract}

\end{abstract}


\maketitle

\section{I. Introduction}

Equilibrium systems, which satisfy detailed balance and therefore do not have any particle or energy current, are based on 
a well founded thermodynamic theory. Studies of equilibrium systems start with the zeroth law which is the corner-stone of 
equilibrium thermodynamics. The zeroth law states that, in equilibrium, there exists a set of intensive variables, each of 
which being conjugate to a corresponding extensive conserved quantity, and these intensive variables equalize when two 
systems are kept in contact and allowed to exchange the conserved quantities. Specifically, an equilibrium system in contact 
with a reservoir is characterized by the familiar Boltzmann distribution where the probability of a microstate $C$ is given by 
$P(C) \sim \exp [-\beta\{ H(C)-\mu N \} ]$ with $H(C)$ the internal energy of configuration $C$, $N$ the number of particles 
in the system, and the intensive variables $\beta$ and $\mu$ being the inverse temperature and the chemical potential of the 
reservoir, respectively. These variables $\beta$ and $\mu$ are conjugate to energy and particle-number of the system. Moreover, 
in equilibrium, there is a class of general fluctuation-response relations, collectively called the 
Fluctuation Dissipation Theorem (FDT), which relate the response of a system upon change of an intensive variable (e.g., 
chemical potential) to the fluctuation in the corresponding extensive variable (e.g., particle-number).

One could inquire whether there can be a similar thermodynamic structure for nonequilibrium systems as well. Among the 
vast class of nonequilibrium systems, one ubiquitous subclass is that of systems having nonequilibrium steady state (NESS)
\cite{Zia_NESS}. The systems in a NESS have time-independent macroscopic properties which is similar to that of systems in 
equilibrium. However, unlike in equilibrium, they have a steady current and generally cannot be characterized by the Boltzmann 
distributions with an a-priori known energy functions. Perhaps not surprisingly, even for this conceptually simplest class of 
driven systems with NESS, there is no well founded thermodynamic structure. Intensive studies in this direction to find a 
suitable framework for description of NESS have been undertaken \cite{Eyink1996, Oono_Paniconi1998, Bertini_etal, 
Bodineau_Derrida, Sasa2006}. In this paper, we ask whether a homogeneously driven many-particle system can be characterized in 
terms of an intensive thermodynamic variable which equalizes for two systems in contact. We address this question using a simple 
class of stochastic models called driven lattice gases.

Although there have been many attempts to define an intensive variable for driven systems in various specific contexts
\cite{Wang_Menon, Henkes, Shokef, Hayashi_Sasa2003}, there was no general formulation in this regard until recently
when a prescription to define such a variable for driven systems was proposed by invoking a hypothesis called the 
asymptotic factorization property \cite{Bertin_PRL2006}. This property has been shown to be satisfied for a class of 
driven systems having short-range spatial correlations such as the zero range process (ZRP).

The ZRP is one of the simplest example of driven interacting many-particle systems which do not satisfy detailed balance and 
have nonequilibrium steady states \cite{Evans}. Previously, the ZRP has been mainly considered as model-system for various mass 
transport processes and has been used to study the phenomenon of condensation transition in nonequilibrium systems 
\cite{Majumdar}. Recently, it has been demonstrated \cite{Bertin_PRE2007} that the systems governed by the ZRP has a simple 
thermodynamic structure where a suitably defined intensive variable, like equilibrium chemical potential, indeed equalizes for 
two such systems in contact. There is also a corresponding fluctuation-response relation between the compressibility and the 
fluctuation in the particle-number, which is satisfied exactly for a system in contact with a particle reservoir. The main 
advantage of studying these simple models like the ZRP was that the steady-state probability distributions, which have simple 
factorized forms, can be calculated exactly and therefore various features of driven systems in contact can be studied 
analytically. Due to the simple form of the steady-state distribution, it was also possible to analyze the role of dynamics at 
the contact between two systems.

However for driven systems with nontrivial steady-state properties, the situation is expected to be far more complex. Here we 
consider a simple model of a driven interacting many-particle system known as Katz-Lebowitz-Spohn (KLS) model \cite{Zia, KLS}. 
The KLS model, first introduced to study fast ionic conductors \cite{Dietrich}, is a paradigm in nonequilibrium statistical 
mechanics. The model describes a stochastic lattice gas of charged particles which is homogeneously driven by a constant 
externally applied electric field. Initially, the primary motivation behind studying the KLS model was to understand the 
nontrivial spatial structures and phase transitions in generic bulk-driven systems with nonequilibrium steady states. Since the 
introduction, the KLS model has been studied intensively. By now, the phase diagram in the plane of temperature and electric 
field strength is quite well known, mainly from extensive simulations \cite{KLS, Zia, review1, Leung-Schmittmann-Zia} as well 
as the results from mean-field theory \cite{review2} and renormalization-group analysis of a continuum version of the model 
\cite{review3, Leung-Schmittmann-Zia}. However there is still no well founded thermodynamic theory for these driven interacting 
many-particle systems.

In this paper, we explore the ``equilibration'' between two driven lattice gases when they are brought into contact. Recently 
we studied the KLS model, with repulsive nearest-neighbor interactions among particles \cite{Pradhan_PRL}, which revealed a 
simple but approximate thermodynamic structure. Here we extend our previous studies to the systems with attractive interactions 
as well. Interestingly, both for attractive as well as for repulsive interactions and for a wide range of parameter values, we 
find 
that, to a very good approximation, there is an intensive thermodynamic variable, like equilibrium chemical potential, which 
determines the final steady state while two systems are allowed to exchange particles. Consequently, the zeroth law of 
thermodynamics as well as the fluctuation-response relation between the compressibility and the fluctuations in particle-number 
are satisfied remarkably well in a wide range of parameter values.

However there are also observable deviations from this simple thermodynamic structure, especially at high interaction strengths 
and large driving fields. We explain these deviations by expressing the asymptotic factorization property, which was 
initially proposed by Bertin {\it et al.} in \cite{Bertin_PRL2006, Bertin_PRE2007} and later discussed by us for driven lattice 
gases in \cite{Pradhan_PRL}, in a modified form where contributions to the large deviation functions due to the contact dynamics 
are identified. To illustrate the origin of these deviations, we study the nontrivial role of the contact dynamics using first a 
simple variant of the ZRP and later the KLS model. We find that, depending on the various parameter values, the contact 
dynamics can amount to an excess chemical potential across the contact for the variant of the ZRP as well as the KLS model. This 
excess chemical potential is generally a-priori unknown, and in some sense arbitrary for an arbitrarily chosen contact dynamics. 
Therefore, it may not be always possible to assign, to the individual systems, an intensive variable which is independent of the 
contact between the two systems, thus accounting for the deviations from the zeroth law.

Here is a brief outline of the paper. In section II we describe the model, in section III we present the numerical results 
concerning the zeroth law for the KLS model and possible deviations from the law. In section IV, we discuss the excess chemical 
potential for a variant of the ZRP as well as a variant of the equilibrium KLS model. In section V, we describe how the large 
deviation principle can be written in a modified form which can capture the deviations from the zeroth law and then
discuss the excess chemical potential for the KLS model. At 
the end, we summarize. In Appendix A, we give proof of the ansatz for the steady-state probability distribution in the 
case of the ZRP and, in Appendix B, we discuss the fluctuation-response relation for the ZRP as well as for the KLS model.

\section{II. Model}

\begin{figure}
\begin{center}
\leavevmode
\includegraphics[width=8.0cm,angle=0]{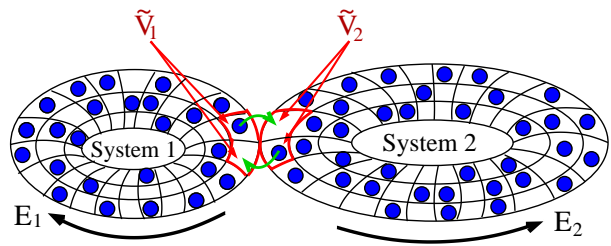}
\caption{A schematic diagram of two nonequilibrium steady states in contact
with the contact region $\tilde{V}_1$ and $\tilde{V}_2$. Particles are allowed to
be exchanged through the contact region $\tilde{V}_1$ and $\tilde{V}_2$. In this process
the total number of particles $N=N_1+N_2$ are conserved where $N_1$ and $N_2$ are the 
number of particles in system 1 and system 2 respectively.}
\label{dia_contact}
\end{center}
\end{figure}

We consider stochastic lattice gases of charged particles which are driven out of equilibrium by constant externally applied 
electric fields in the bulk and which therefore have spatially homogeneous steady states \cite{KLS}. Particles move on a 
discrete lattice and jump stochastically from one site to any of its nearest-neighbor sites, preferably towards the direction 
of the external driving field of magnitude $E$. Due to hardcore repulsion among particles, a lattice site can be occupied by 
at most one particle. In addition, particles may also interact with each other through a nearest-neighbor pair-potential of 
interaction strength $K$. We define the occupation variable $\eta({\bf r})$ at a site ${\bf r} \equiv \{r_x,r_y\}$ where 
$\eta({\bf r})=0,1$: if a site ${\bf r}$ is occupied $\eta({\bf r})=1$ otherwise $\eta({\bf r})=0$. We consider two systems 
of lattice gases where system 1 and system 2 are confined respectively in volume $V_1$ and volume $V_2$ (see Fig. 
\ref{dia_contact}). When two such systems are brought into contact, they are connected 
at a finite set of points $V_1'$ and $V_2'$, subsets of $V_1$ and $V_2$ respectively ($V_1', V_2' \ll V_1, V_2$) and, while 
in contact, they can exchange particles with each other. The energy function $H$ of the combined two systems is given by 
\bea H = K_1 \sum_{\langle {\bf r_1}, {\bf r_1'} \rangle} \eta({\bf r_1}) \eta({\bf r_1'}) 
+ K_2 \sum_{\langle {\bf r}_2, {\bf r}_2' \rangle} \eta({\bf r}_2) \eta({\bf r}_2')
\label{H1} 
\eea 
where $\langle *,* \rangle$ denotes sum over nearest-neighbor sites with ${\bf r_1}, {\bf r_1}' \in V_1$ and ${\bf r_2}, 
{\bf r_2}' \in V_2$, $K_1$ and $K_2$ strength of interactions among particles for the respective systems. Note that systems 
may in general have different microscopic dynamics depending on the interaction strengths $K_1$ and $K_2$. Constant 
driving fields, $E_1$ and $E_2$ respectively in systems 1 and 2, are applied along $x$-directions, with periodic boundary 
conditions imposed in both $x$ and $y$ directions. Driven bilayer systems have been considered before where particles can 
jump from one layer to the other at any site \cite{Bilayr_systems}. However here we consider the case where particles can 
jump from one system to the other through a very small contact region between the systems.

We choose jump rates of particles such that they satisfy the local detailed balance condition \cite{KLS}. A pair of nearest
-neighbor sites, located at ${\bf r}$ and ${\bf r'}$, in a configuration $C$ are chosen randomly and an attempt is made to
interchange the occupation variables where the attempted final configuration is denoted by $C'$. Let us denote the 
corresponding transition rate from configuration $C$ to $C'$ as $w(C'|C)$. For movements of particles inside the same system
(i.e., particles not jumping from one system to the other), a quantity $\Delta(E) = H(C') - H(C) - E(r_x-r_x')$, which depends 
on the driving field $E$, is defined where $r_x$ denotes the $x$-component of the position vector ${\bf r}$ and $H(C)$ is the 
energy of the configuration $C$. The transition rate is assigned to be 
\be 
w(C'|C)=\mbox{min} \{1, e^{-\beta \Delta(E)}\}
\label{Metropolis1}
\ee
where $\beta$ is the inverse temperature of the heat bath. When the chosen pair of sites are such that a jump is attempted 
from one system to the other across the contact, the transition rate is assigned to be 
\be 
w(C'|C)=\mbox{min} \{1, e^{-\beta \Delta(0)}\}
\label{Metropolis2}
\ee
where $\Delta(0) = H(C') - H(C)$. Note that there is no field along the bonds connecting the two systems. In the 
simulations, we consider two-dimensional systems ($V=L \times L$) with periodic boundaries in both directions. We put 
$\beta =1$ throughout the paper.

When $E_1=E_2=0$, the jump rates satisfy the detailed balance condition and the configuration $C$ of the combined system has the 
Boltzmann probability distribution $P(C) \sim \exp[- H(C)]$. For $E_1, E_2 \ne 0$, there is a constant current in the steady 
state. However, unlike in equilibrium, the nonequilibrium steady-state probability distribution in general is not given by the 
Boltzmann
distribution with an a-priori known energy function and, except for a few cases (e.g., with only hard-core interactions), the 
steady-state probability distribution is not known. When two systems are brought into contact, they can exchange particles with 
the total number of particles $N=N_1+N_2$ conserved where $N_1=\sum_{{\bf r_1} \in V_1} \eta({\bf r_1})$ and $N_2=\sum_{{\bf r_2} 
\in V_2} \eta({\bf r_2})$ are number of particles in system 1 and system 2, respectively. In the sections below, we consider the 
cases where the conserved quantity is the number of particles and therefore attempt to define an intensive variable, called 
chemical potential in analogy with equilibrium, which is conjugate to the conserved particle-number.

\section{III. Numerical Results}

\subsection{A. The Zeroth Law}

\begin{figure}[h!]
\begin{center}
\includegraphics [scale=0.2] {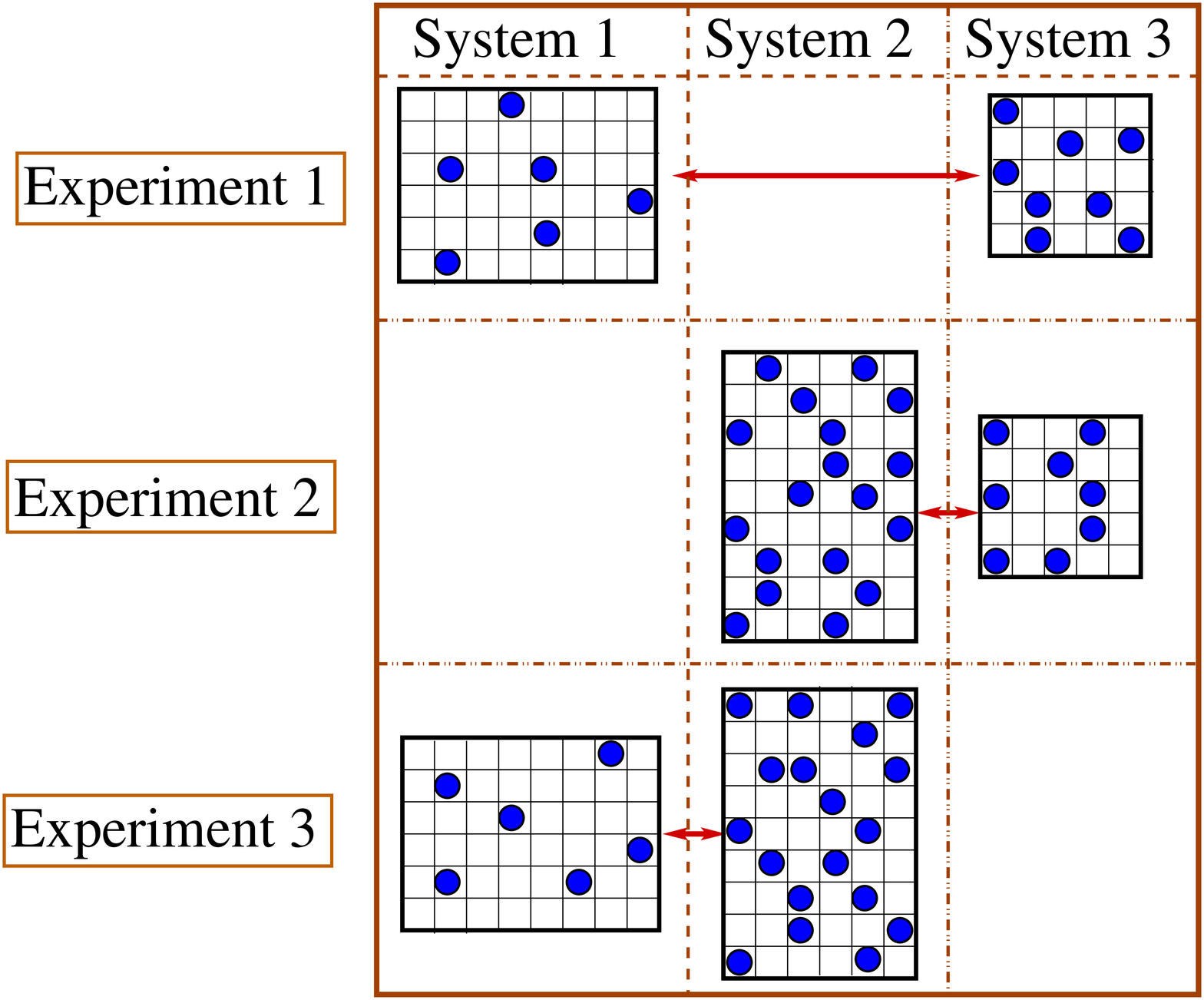}
\caption{Schematic diagram to test the zeroth law of thermodynamics for equilibrium systems using a sequence 
of three thought-experiments where three systems (for simplicity, shown in two dimensions) are separately kept 
in contact with each other and allowed to equilibrate. Experiment 1: System 1 and system 3 in contact.
Experiment 2: System 2 and system 3 in contact. Experiment 3: System 1 and system 2 in contact.}
\label{Dia_zeroth_law}
\end{center}
\end{figure}

First let us describe the zeroth law in the context of equilibrium systems. We consider three systems and perform the following 
three thought-experiments which are schematically presented in Fig. \ref{Dia_zeroth_law}. In the first experiment, system 1 and 
system 3 are brought into contact and allowed to exchange particles with each other. System 1 and system 3 eventually equilibrate 
and reach a final equilibrium state with constant average densities $n_1$ and $n_3$ respectively. In the second experiment, 
system 2 and system 3 are separately brought into contact and allowed to exchange particles. In this case, the initial density 
of system 2 is tuned such that system 3 has the same final density as that of system 3 in the first experiment. Let us denote
the final equilibrium densities for system 2 and 3 in the second experiment as $n_2$ and $n_3$ respectively. Now in the 
third experiment, system 1 and system 2 with initial densities $n_1$ and $n_2$ respectively are brought into contact and allowed 
to exchange particles. One could ask what the final densities in this case would be. The zeroth law of thermodynamics provides 
the answer that the final densities will not change any more and will be exactly equal to the respective initial densities. 
Thus the zeroth law allows us to assign to an equilibrium system an intensive thermodynamic variable called the chemical 
potential which equalizes for two systems in contact. Note that, in Fig. \ref{Dia_zeroth_law} if one compares the density 
profiles of two systems being in the same column, the corresponding average density profiles would be exactly same. 

\begin{figure}[h!]
\begin{center}
\includegraphics [scale=0.70] {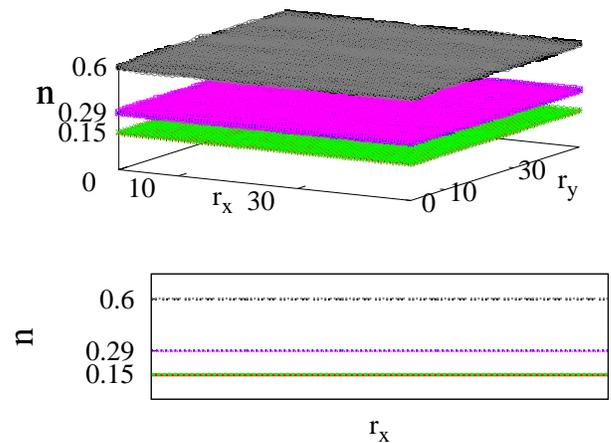}
\caption{Numerical experiments to test the zeroth law. Top panel: Average densities as a function of position co-ordinates 
$r_x$ and $r_y$. Bottom panel: Cross-sections (along $x$-direction) of the density profiles.  
Driven system 1: $K=-1$, $E=6$ (top density profiles). Driven system 2: $K=-0.75$, $E=4$ 
(middle density profiles). Equilibrium system 3: $K=0$, $E=0$ (bottom density profiles). 
All systems considered here are with same volume  $V=50 \times 50$. Experiment 1: System 1 with density $n_1 \simeq 0.60$ 
(top grey profile) ``equilibrated'' with system 3 with density $n_3 \simeq 0.15$ (bottom red profile). Experiment 2: System 2 
with density $n_2 \simeq 0.29$ (middle magenta profile) ``equilibrated'' with system 3 with density $n_3 \simeq 0.15$ 
(bottom green profile). Experiment 3: System 1 with density $n_1'$ (top black profile) ``equilibrated'' with system 2 with 
density $n_2'$ (middle blue profile) where $n_1' \simeq n_1$ and $n_2' \simeq n_2$. }
\label{zeroth_law_attractive}
\end{center}
\end{figure}

\begin{figure}[h!]
\begin{center}
\includegraphics [scale=0.70] {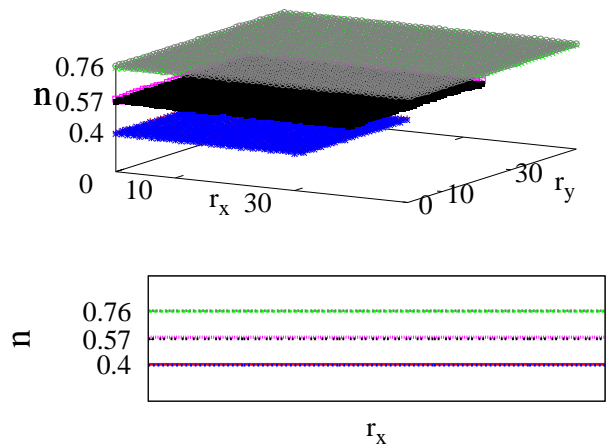}
\caption{Numerical experiments to test the zeroth law. Top panel: Average densities as a function of position co-ordinates 
$r_x$ and $r_y$. Bottom panel: Cross-sections (along $x$-direction) of the density profiles.
Driven system 1: $K=3.75$, $E=6$ and $V=32 \times 32$ (bottom density profiles). 
Driven system 2: $K=1.5$, $E=5$ and $V=40 \times 40$ (middle density profiles). 
Equilibrium system 3: $K=0.75$, $E=0$ and $V=50 \times 50$ (top density profiles). 
Experiment 1: System 1 with density $n_1 \simeq 0.40$ (bottom red profile) ``equilibrated'' with system 3 with density 
$n_3 \simeq 0.76$ (top grey profile). Experiment 2: System 2 with density $n_2 \simeq 0.57$ (middle black profile) 
``equilibrated'' with system 3 with density $n_3 \simeq 0.76$ (top green profile). Experiment 3: System 1 with density 
$n_1'$ (bottom blue profile) ``equilibrated'' with system 2 with density $n_2'$ (middle magenta profile) where 
$n_1' \simeq n_1$ and $n_2' \simeq n_2$. }
\label{zeroth_law_repulsive}
\end{center}
\end{figure}

We use the above strategy to test the zeroth law for systems in nonequilibrium steady states. We perform the same set of 
numerical experiments as described above, but now with two of the systems driven out of equilibrium due to external driving 
fields present in the bulk of the individual systems. Interestingly, similar to the equilibrium case, we observe that, for 
various values of interaction strengths, driving fields and densities, the zeroth law is quite well satisfied, i.e, if two 
driven lattice gases are separately  ``equilibrated'' with a common system with a fixed density, they will also equilibrate 
amongst themselves. Two such examples are given below.
\\
\\
{\it 1. Example for driven systems with attractive interactions} -  In Fig. \ref{zeroth_law_attractive} we consider three 
systems, a driven system 1 with $K=-1$, $E=6$, a driven system 2 with $K=-0.75$, $E=4$ and an equilibrium system 3 with $K=0$, 
$E=0$. First, system 1 with density $n_1 \simeq 0.60$ and system 2 with density $n_2 \simeq 0.29$ are separately 
``equilibrated'' with system 3 with a fixed density $n_3 \simeq 0.15$. We then find that system 1 and system 2, with the 
initial densities $n_1 \simeq 0.60$ and $n_2 \simeq 0.29$ respectively, ``equilibrate'' with each other such that, to a very 
good approximation, the respective final steady-state values of densities $n_1' \simeq 0.60$ and $n_2' \simeq 0.29$ remain 
almost unchanged. 
\\
\\
{\it 2. Example for driven systems with repulsive interactions} - In Fig. \ref{zeroth_law_repulsive}, we consider three systems, 
a driven system 1 with $K=3.75$, $E=6$, a driven system 2 with $K=1.5$, $E=5$ and an equilibrium system 3 with $K=0.75$, $E=0$. 
System 1 with density $n_1 \simeq 0.40$ and system 2 with density $n_2 \simeq 0.57$ are separately ``equilibrated'' with system 
3 with a fixed density $n_3 \simeq 0.76$. In Fig. \ref{zeroth_law_repulsive}, one could see that system 1 and system 2 with the 
above initial densities $n_1 \simeq 0.40$ and $n_2 \simeq 0.57$ respectively equilibrate with each other with the respective 
final densities $n_1' \simeq 0.39$ and $n_2' \simeq 0.58$. Also in this case, the zeroth law is satisfied to a good approximation 
where final densities $n_1'$ and $n_2'$ remain almost same as the initial densities.

\begin{figure}[h!]
\begin{center}
\includegraphics [scale=0.70] {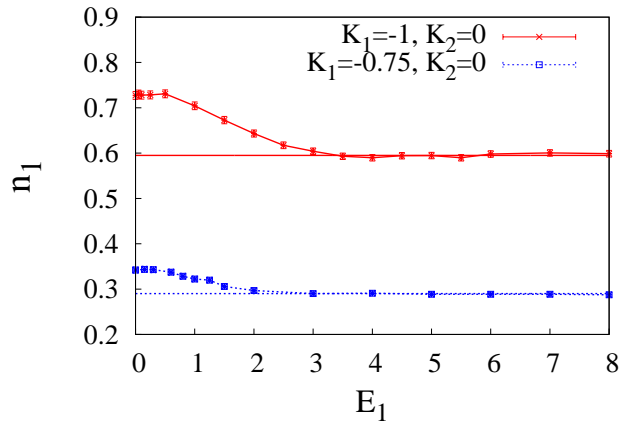}
\includegraphics [scale=0.70] {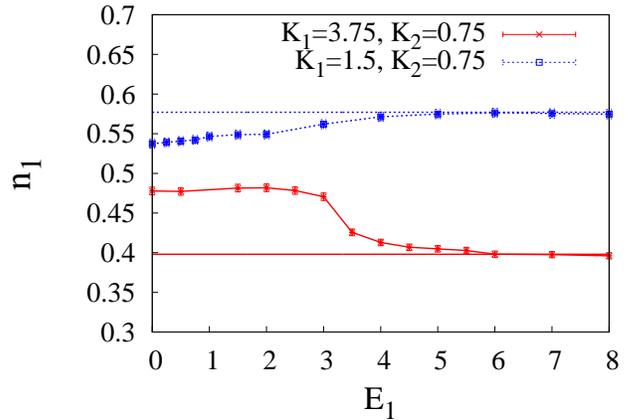}
\caption{Density $n_1$ of driven system 1 as a function of driving field $E_1$ when the system 1 is in contact with an 
equilibrium system 2 
with a fixed density $n_2$. Top panel: Densities of two driven systems with interaction strengths $K_1=-1$ and $K_1=-0.75$ 
plotted as a function of respective driving fields $E_1$ when the systems are separately in contact with an equilibrium system 
with 
$K_2=0$ and $n_2 \simeq 0.15$. Bottom panel: Densities of two driven systems with interaction strengths $K_1=3.75$ and $K_1=1.5$ 
plotted as a function of respective driving fields $E_1$ when the systems are separately in contact with an equilibrium system 
with $K_2=0.75$ and $n_2 \simeq 0.76$.  }
\label{n_vs_E1}
\end{center}
\end{figure}

Interestingly, as seen in Figs. \ref{zeroth_law_attractive} and \ref{zeroth_law_repulsive}, two systems in contact have 
homogeneous density profiles even when one or both of them may be driven. The driven systems are indeed far away from equilibrium 
since the numerical values of currents in the nonequilibrium steady states considered in Figs. \ref{zeroth_law_attractive} and 
\ref{zeroth_law_repulsive} are near to the respective maximum values of currents (data not shown). Moreover in the top panel of 
Fig. \ref{zeroth_law_repulsive}, system 1 with density $n_1 \simeq 0.40$ has a homogeneous disordered state in contrast to an 
ordered state for the corresponding equilibrium system. The equilibrium system, with the same interaction strength $K=3.75$ and 
the same density, has a symmetry-broken ordered phase with a checkerboard-like pattern where sub-lattice densities are different 
\cite{Zia}.

Importantly, the macroscopic properties like densities do indeed depend on the driving field when a driven system 1 is kept in 
contact with an equilibrium system 2 with a fixed density $n_2$. This can be seen in the behavior of density $n_1$ of a driven 
system as a function of driving field $E_1$. We consider the driven systems which were previously used to test the zeroth law in 
Figs. \ref{zeroth_law_attractive} and \ref{zeroth_law_repulsive}. In the top panel of Fig. \ref{n_vs_E1}, densities of two driven 
systems with attractive interactions, with $K_1=-1$ and $K_1=-0.75$, have been plotted as a function of the driving fields $E_1$ 
in the respective systems where both the systems are separately kept in contact with an equilibrium system with $K_2=0$ and a 
fixed density $n_2 \simeq 0.15$. In the bottom panel of Fig. \ref{n_vs_E1}, densities of two other driven systems with repulsive 
interactions, with $K_1=3.75$ and $K_1=1.5$, have been plotted as a function of the driving fields $E_1$ in the respective 
systems where both the systems are now separately kept in contact with an equilibrium system with $K_2=0.75$ and a fixed density 
$n_2 \simeq 0.76$. One could see that the densities vary quite significantly, almost by an amount of $10\%$ or more from the 
respective equilibrium values depending on the interaction strengths, when the driving field varies from $E_1=0$ to a large 
value $E_1 \gg K_1$.

\begin{figure}[h!]
\begin{center}
\includegraphics [scale=0.70] {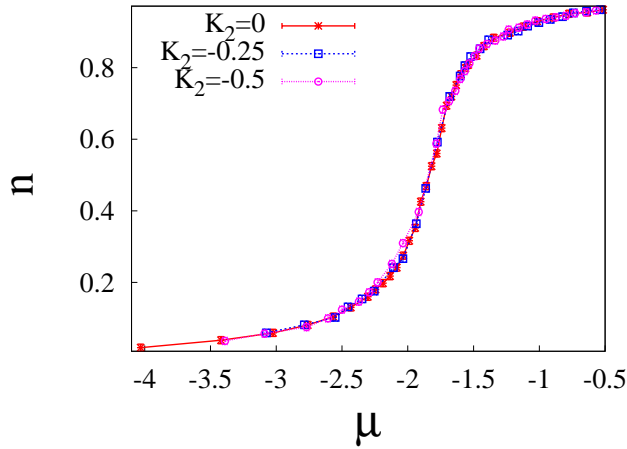}
\caption{Density $n$ {\it vs.} chemical potential $\mu$ is plotted for a driven system with $K_1=-1$, $E_1=6$ 
which is separately kept in contact with three equilibrium systems with interaction strengths $K_2=0$, 
$K_2=-0.25$, and $K_2=-0.5$. The collapse of the curves agrees well with the zeroth law.}
\label{0thLaw1}
\end{center}
\end{figure}

Provided that the zeroth law is satisfied, one can define a chemical potential even for a driven system as following. 
A driven system is kept in contact with an equilibrium one and allowed to reach a steady state. Then, in the steady state, 
one can simply assign the chemical potential of the equilibrium system to the driven one. For nonzero interaction strengths 
$K \ne 0$, even the equilibrium chemical potential $\mu$ cannot be calculated directly as one does not know the explicit form 
of $\mu (n)$ as a function of density $n$. However, for equilibrium system with noninteracting hardcore particles $K=0$, 
the chemical potential $\mu$ can be expressed as a function of density $n$ where
\be
\mu = \ln \left( \frac{n}{1-n} \right),
\label{mu}
\ee 
by using the relation $\mu=-(\partial s/\partial n)$ where $s=-[n \ln n + (1-n) \ln (1-n)]$ is the equilibrium entropy per 
lattice site. Since the zeroth law is exactly satisfied in equilibrium, the chemical potential for any equilibrium system 
with $K\ne 0$ can be measured by keeping the system in contact with an equilibrium system with $K=0$ and then assigning a 
chemical potential for a system with $K=0$ to that with $K\ne0$. In Fig. \ref{0thLaw1}, we plot the density $n$ as a function 
of chemical potential $\mu$ when a driven system with $K_1=-1$, $E_1=6$ is separately kept in contact with three equilibrium 
systems
with various interaction strengths $K_2=0$, $K_2=-0.25$, and $K_2=-0.5$. Given that the zeroth law is satisfied, the various 
curves 
for density as a function of chemical potential should fall on each other. The collapse of curves in Fig. \ref{0thLaw1}
indeed agrees quite well with the zeroth law.

\begin{figure}[h!]
\begin{center}
\includegraphics [scale=0.70] {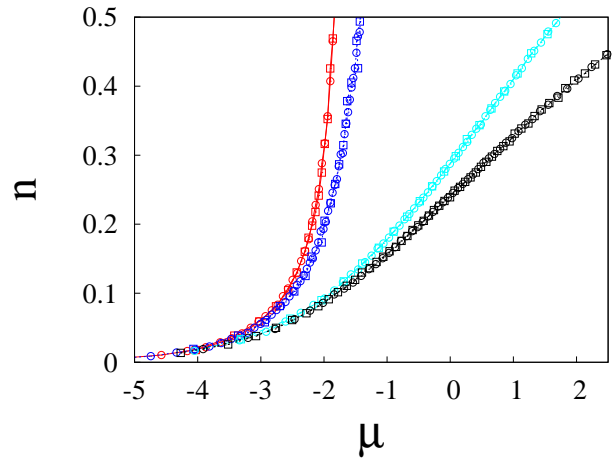}
\caption{Density $n$ as a function of chemical potential $\mu$ for driven systems (fixed $K_1$, $E_1$) with different 
values of system size $V_1$ in contact with an equilibrium system of noninteracting 
hardcore particles ($K_2=0, E_2=0$) with different values of system size $V_2$.
Circles correspond to the case when the system 1 with $V_1=32\times 32$ is in contact with system 2 with $V_2=32\times 32$.
Squares correspond to the case when the system 1 with $V_1=20\times 20$ is in contact with system 2 with $V_2=100\times 100$.
For the curves from left to right: (1) two curves (red) for systems with $K_1=-1$, $E_1=6$, 
(2) two curves  (blue) for systems with $K_1=-0.75$, $E_1=4$, (3) two curves  (sky blue) for systems 
with $K_1=1$, $E_1=6$, and (4) two curves  (black) for systems with $K_1=2$, $E_1=2$.}
\label{n_vs_mu_systemsizes1}
\end{center}
\end{figure}

We have studied the dependence of densities on the system sizes as well. We find that the densities of two systems in contact 
with each other, when one or both systems may be driven, are almost independent of system sizes. In Fig. 
\ref{n_vs_mu_systemsizes1} we plot the density $n$ as a function of chemical potential $\mu$ for various system sizes, and for 
various values of interaction strengths and driving fields. The densities seem to only 
depend on the interaction strengths and driving fields of the respective systems.

\begin{figure}[h!]
\begin{center}
\includegraphics [scale=0.70] {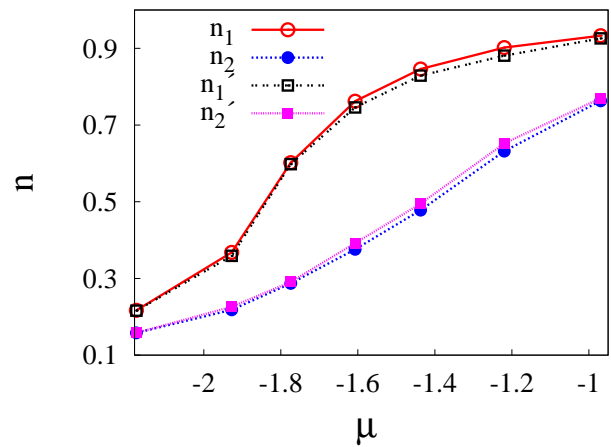}
\caption{Final densities $n_1'$ and $n_2'$ for two driven systems, respectively with $K=-1$, $E=6$ and $K=-0.75$, $E=4$, 
being in contact with each other are compared with their respective initial densities $n_1$ and $n_2$. The initial densities 
correspond to those obtained by keeping the driven systems separately in contact with an equilibrium system of a fixed chemical 
potential $\mu$. If the zeroth law is satisfied, the initial and final densities should be exactly the same for a given $\mu$. 
}
\label{0thLaw_density1}
\end{center}
\end{figure}

\begin{figure}[h!]
\begin{center}
\includegraphics [scale=0.67] {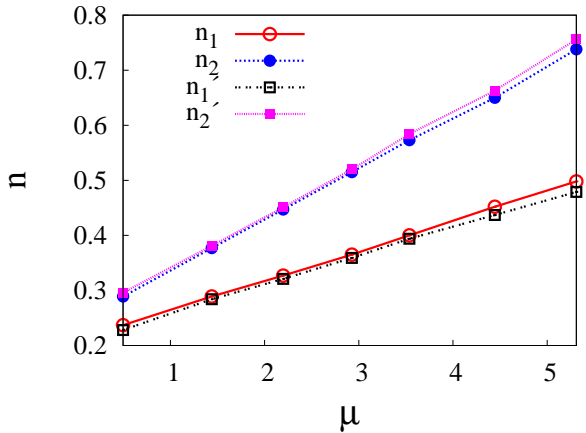}
\caption{Final densities $n_1'$ and $n_2'$ for two driven systems, respectively with $K=3.75$, $E=6$ and $K=1.5$, $E=5$, 
being in contact with each other are compared with their respective initial densities $n_1$ and $n_2$. The initial densities 
correspond to those obtained by keeping the driven systems separately in contact with an equilibrium system of a fixed chemical 
potential $\mu$. If the zeroth law is satisfied, the initial and final densities should be exactly the same for a given 
$\mu$.}
\label{0thLaw_density2}
\end{center}
\end{figure}

In Figs. \ref{zeroth_law_attractive} and \ref{zeroth_law_repulsive}, we have tested the zeroth law only for a particular set 
of values of densities. We now study the zeroth law for various other densities where the interaction strengths and driving 
fields are kept fixed. We consider driven systems 1 and 2 with respective densities $n_1$ and $n_2$ where the systems are 
separately equilibrated with an equilibrium system 3 with density $n_3$ and corresponding chemical potential $\mu$. Then system 
1 and system 2 with the respective initial densities $n_1$ and $n_2$ are equilibrated with each other where they eventually 
reach final densities $n_1'$ and $n_2'$. Clearly, if the zeroth law is satisfied, then the corresponding initial and final 
densities would be exactly the same, i.e., $n_1=n_1'$ and $n_2=n_2'$. We consider the systems as previously discussed in 
Figs. \ref{zeroth_law_attractive} and \ref{zeroth_law_repulsive}. In Fig. \ref{0thLaw_density1}, we plot $n_1$ and $n_1'$ for 
system 1 with $K=-1$, $E=6$ and $n_2$ and $n_2'$ for system 2 with $K=-0.75$, $E=4$ as a function of chemical potential $\mu$ 
of the equilibrium system 3, with $K=E=0$. Similarly, in Fig. \ref{0thLaw_density2}, we plot $n_1$ and $n_1'$ for system 1 with 
$K=3.75$, $E=6$ and $n_2$ and $n_2'$ for system 2 with $K=1.5$, $E=5$ as a function of chemical potential $\mu$ of the system 3, 
with $K=0.75$, $E=0$. Although, the zeroth law is satisfied to a good approximation, there are indeed small but observable 
deviations from the law, i.e., up to $5\%$ deviations in the final densities from the corresponding initial density values. 
In the next section we discuss the deviations from the zeroth law in more detail.

\subsection{B. Deviations from the Zeroth Law}

\begin{figure}[h!]
\begin{center}
\includegraphics [scale=0.65] {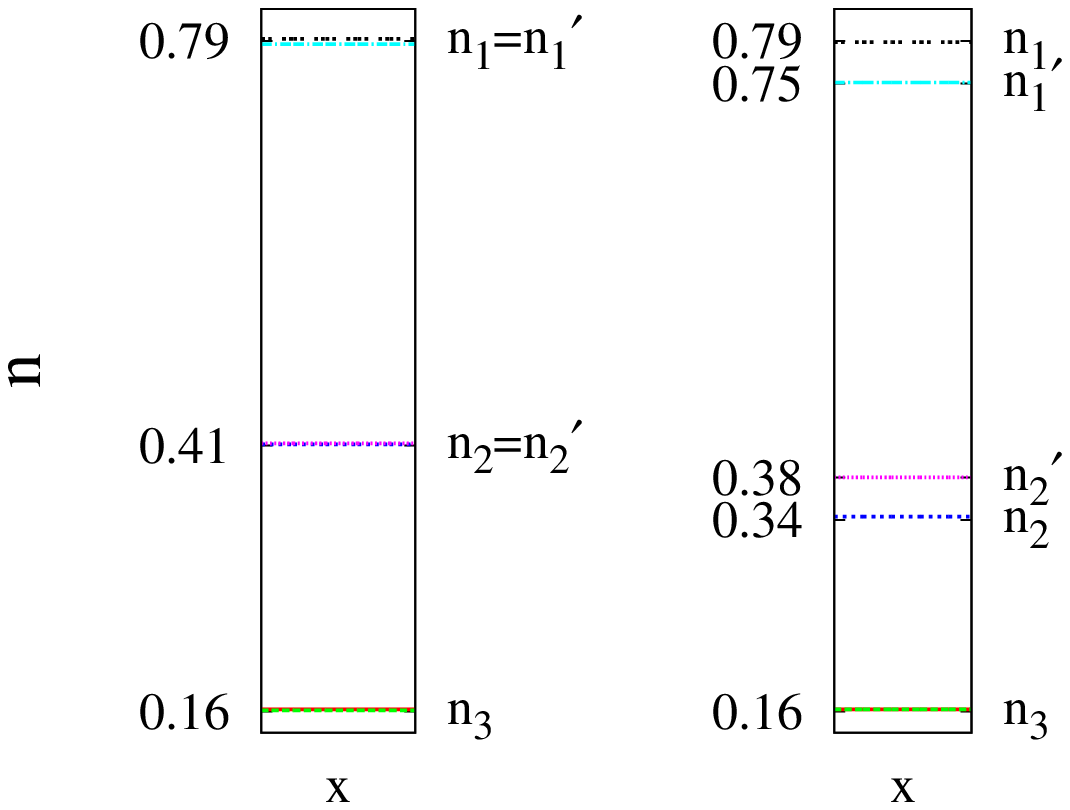}
\includegraphics [scale=0.65] {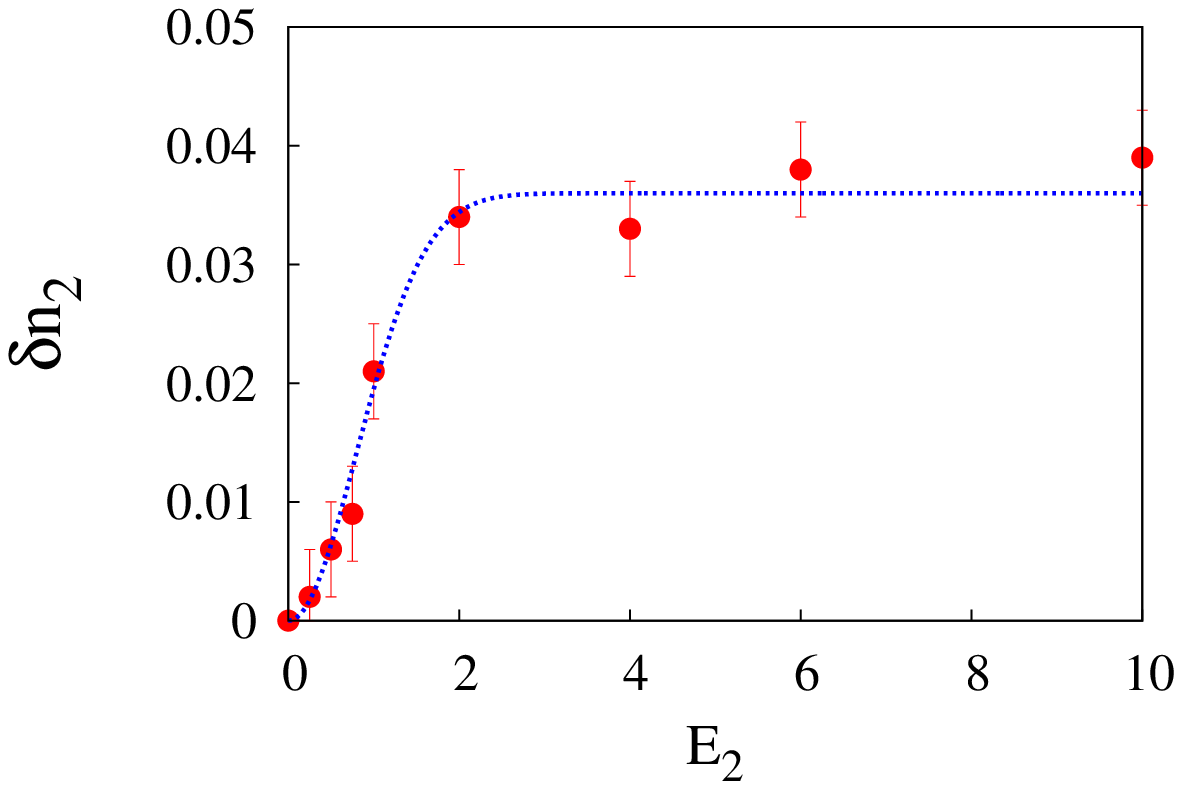}
\caption{ In top panel, the cross-sections (along $x$-direction) of the density profiles are plotted. Systems 1, 2 and 3 (all 
with same size $V=50 \times 50$) have interaction strengths $K=-1$ (top density profiles), $K=-0.75$ (middle density profiles) 
and $K=0$ (bottom density profiles) respectively. First, system 1 with density $n_1$ and system 2 with density $n_2$ are 
separately equilibrated with system 3 with density $n_3$. Then system 1 and 2, with initial densities $n_1$ and $n_2$ 
respectively, are equilibrated with each other where the final steady-state densities are $n_1'$ and $n_2'$ respectively. Top 
panel (left): For equilibrium systems where $n_1=n_1'$ and $n_2=n_2'$. Top panel (right): For the case when system 2 driven 
with field $E_2=10$ with $n_1 \ne n_1'$ and $n_2 \ne n_2'$. Bottom panel: Difference in density $\delta n_2 = n_2'-n_2$ 
is plotted versus driving field $E_2$ for system 2. The blue line, which is a fitting function $(a-b e^{-\kappa E_2^2})$ with
$a=0.036$, $b=0.036$ and $\kappa = 0.78$, is a guide to the eye.
}
\label{zeroth_law_violation1}
\end{center}
\end{figure}

In the previous section, we have seen that, for a large parameter range, the driven lattice gases have an approximate 
but remarkably simple thermodynamic structure. However, it should be noted that there are also larger observable 
deviations from the simple thermodynamic law as discussed next.

In Fig. \ref{zeroth_law_violation1}, we perform numerical experiments similar to that discussed before (e.g., see Figs. 
\ref{zeroth_law_attractive} and \ref{zeroth_law_repulsive}) where system 1 with density $n_1$ and system 2 with density $n_2$ 
are separately equilibrated with system 3 with a fixed density $n_3$. Then system 1 and system 2, with the respective initial 
densities $n_1$ and $n_2$, are equilibrated with each other where the final steady-state densities are $n_1'$ and 
$n_2'$ respectively. In the top-left panel of Fig. \ref{zeroth_law_violation1}, six density profiles along $x$-direction are 
plotted for equilibrium systems 1, 2, and 3. Expectedly, for equilibrium systems, the zeroth law is exactly satisfied where 
$n_1=n_1' \simeq 0.79$ (top density profiles) and $n_2=n_2' \simeq 0.41$ (middle density profiles), i.e., initial densities 
are equal to the respective final densities. In the top-right panel of Fig. \ref{zeroth_law_violation1}, six density profiles 
along $x$-direction are plotted when systems 1 and 3 are equilibrium systems but system 2 is driven with a large field 
$E_2=10$. In this case, the zeroth law is observed to be violated significantly where $n_1 \simeq 0.79 \ne n_1' \simeq 0.75$ 
(top density profiles) and $n_2 \simeq 0.34 \ne n_2' \simeq 0.38$ (middle density profiles), i.e., final densities change 
appreciably as compared to the respective initial densities. In the bottom panel of Fig. \ref{zeroth_law_violation1}, the 
density difference $\delta n_2 = n_2' - n_2$ is plotted as a function of the driving field $E_2$ in system 2.

\begin{figure}[h!]
\begin{center}
\includegraphics [scale=0.70] {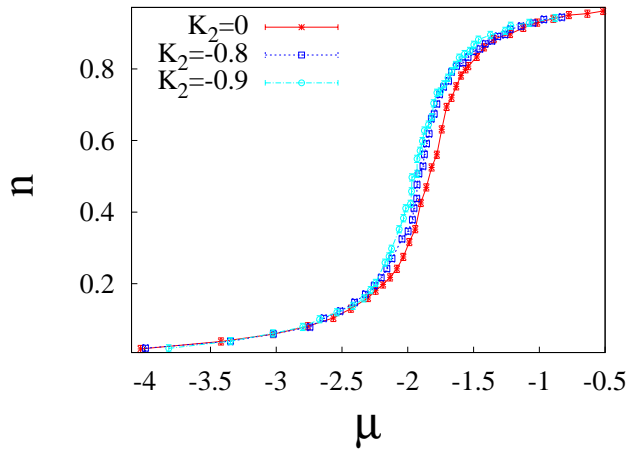}
\caption{Density $n$ as a function of chemical potential $\mu$ for
a driven system with $K_1=-1$, $E_1=6$ which is separately kept in contact with 
three equilibrium systems with interaction strengths $K_2=0$, $K_2=-0.8$, and $K_2=-0.9$.}
\label{0thLaw_vio1}
\end{center}
\end{figure}

\begin{figure}[h!]
\begin{center}
\includegraphics [scale=0.70] {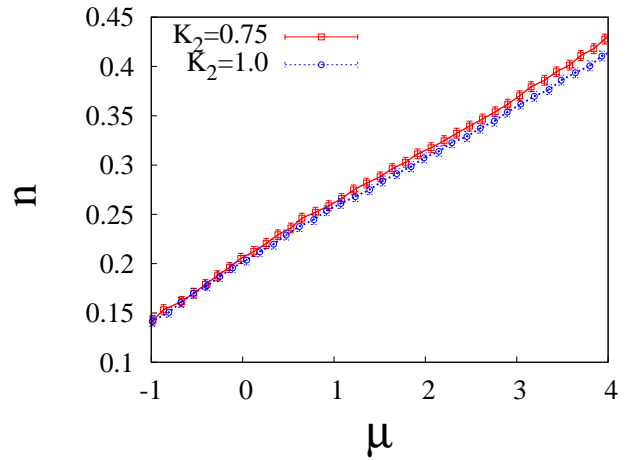}
\caption{Density $n$ as a function of chemical potential $\mu$ for
a driven system with $K_1=3.75$, $E_1=6$ which is separately kept in contact with 
two equilibrium systems with interaction strengths $K_2=0.75$ and $K_2=1$. }
\label{0thLaw_vio2}
\end{center}
\end{figure}

One can also observe the deviations from the zeroth law which is now studied in a slightly different way as follows. 
We first try to assign a chemical potential to a driven system by keeping the system separately in contact with various 
equilibrium systems and then compare density {\it vs.} chemical potential curves. 
In Fig. \ref{0thLaw_vio1}, we plot density $n$ as a function of chemical potential $\mu$ for a 
driven system, with $K_1=-1$, $E_1=6$, which is separately kept in contact with three equilibrium systems
with the following interaction strengths: $K_2=0$, $K_2=-0.8$ and $K_2=-0.9$.
If the zeroth law was satisfied, all the curves should fall on each other. However, we observe that
there are significant deviations which occur especially around density $n=1/2$. In Fig. \ref{0thLaw_vio2},
we plot $n$ {\it vs.} $\mu$ for another driven system, with repulsive interaction $K_1=3.75$, $E_1=6$, being
separately in contact with equilibrium systems with $K_2=0.75$ and $K_2=1$. We again see deviations from the zeroth
law when the density approaches $n=1/2$.

\section{IV. Role of contact dynamics: Exact results}

\subsection{A. Excess Chemical Potential in the ZRP}

We now study a simple class of driven systems, known as the zero range process (ZRP) \cite{Evans}, to understand the role 
of contact dynamics in the context of equalization of thermodynamic variables for driven systems. The ZRP is defined on a 
discrete 
lattice where, unlike the KLS model, there is no restriction on the occupation number of a site, i.e., 
the lattice sites can be occupied by more than one particle. The jump rate of a particle out of any 
site is assumed to depend on the number of particles on the site.

For simplicity, we consider two one-dimensional rings, ring 1 and ring 2, consisting of $L_1$ and 
$L_2$ sites respectively. The rings are kept in contact with each other so that they can exchange particles
through the contact area. In the $\alpha$th ring ($\alpha=1,2$), any site $i_{\alpha}$ 
($i_{\alpha}=1,2, \dots , L_{\alpha}$) is occupied with $n_{i_{\alpha}}$ particles. Any 
configuration $C$ can be specified, by using the occupation numbers at all sites of two rings, 
$C \equiv (\{ n_{i_1} \}, \{ n_{i_2} \})$. Two rings are connected at two sites, 
say $i_1=1$ and $i_2=1$, through which particles can be exchanged between the rings. The dynamics is 
defined as follows: a particle at site $i_{\alpha}$ can jump only in one direction, say 
to the nearest neighbor site $i_{\alpha}+1$ in the clockwise direction (therefore violating detailed balance), 
with rate 
\be
u_{\alpha}(n_{i_{\alpha}}) = v_{\alpha} 
\frac{f_{\alpha}(n_{i_{\alpha}}-1)}{f_{\alpha}(n_{i_{\alpha}})} 
\mbox{~~with~~} \alpha=1,2 
\label{jump_rate}
\ee 
where $f_{\alpha}(n_{i_{\alpha}})$ is a function of occupation number $n_{i_{\alpha}}$ and the factor $v_{\alpha}$ is 
independent of $n_{i_{\alpha}}$. The form of the jump rates is 
the same as given in Eq. \ref{jump_rate} irrespective of jumps in the bulk or jumps from one ring to the other. But the
factors $v_{\alpha}$ can in general be different, i.e., $v_{\alpha} = v^{(b)}_{\alpha}$ when a particle jumps in the bulk of the 
$\alpha$th ring and $v_{\alpha} = v^{(c)}_{\alpha}$ when a particle jumps at the contact from $\alpha$th ring to the other 
ring. For the zero range process, the properties of nonequilibrium steady states  while two systems are in contact has 
been previously studied, but only for a special case when $v^{(b)}_{\alpha}=v^{(c)}_{\alpha}$ and $v^{(c)}_1 = v^{(c)}_2$
\cite{Bertin_PRE2007}. 
Here we consider the more general case when $v^{(b)}_{\alpha} \ne v^{(c)}_{\alpha}$ as well as $v^{(c)}_1 \ne v^{(c)}_2$,
i.e., the factor $v_{\alpha}$ taking different values at the bulk and the contact of the two rings. However, it 
would be easy to see that the factor $v^{(b)}_{\alpha}$ taking different values in two different rings does not change the 
steady-state properties. Therefore, now onwards, we put $v^{(b)}_1 = v^{(b)}_2=1$. Also, for notational simplicity, we 
denote $v_1 \equiv v^{(c)}_1$ and $v_2 \equiv v^{(c)}_2$. For completeness, let us first discuss the special case when 
$v_1=v_2$ \cite{Bertin_PRE2007}. The steady state probability distribution can be written in a factorized form
\bea
P(\{ n_{i_1}\}, \{ n_{i_2} \}) = \frac{1}{Z_N} \left[ \prod_{i_1=1}^{L_1} f_1(n_{i_1}) 
\prod_{i_2=1}^{L_2} f_2(n_{i_2}) \right] \nonumber \\
\times \delta(N_1+N_2-N),
\eea
where $N_1=\sum_{i_1=1}^{L_1} n_{i_1}$ and $N_2=\sum_{i_2=1}^{L_2} n_{i_2}$ are the number of particles in ring 1 and 
ring 2 respectively, $Z_N$ is the normalization constant. The delta function in the above equation takes into account that the 
total number of particles $N=N_1+N_2$ is conserved. Clearly in this case, the joint probability distribution $P(N_1, N_2)$ of 
particle numbers $N_1$ and $N_2$ can be written in a product form 
\be
P(N_1, N_2) = \frac{Z_1(N_1) Z_2(N_2)}{Z_N}
\label{exact_prod_form}
\ee 
where $Z_{\alpha}(N_{\alpha})=\sum_{\{ n_{i_{\alpha}} \}} \prod_{i_{\alpha}=1}^{L_{\alpha}} f_{\alpha}(n_{i_{\alpha}}) 
\delta(\sum_{i_{\alpha}=1}^{L_{\alpha}} n_{i_{\alpha}}-N_{\alpha})$. As discussed in \cite{Bertin_PRL2006, Bertin_PRE2007}, 
in this case, one can define an intensive variable $\mu_{\alpha} = - \partial \ln Z_{\alpha}/\partial 
N_{\alpha}$ which equalizes when two rings are kept in contact, i.e., $\mu_1=\mu_2$.

Now we discuss the general case when $v_1 \ne v_2$ which leads to our main points. Interestingly, as shown in Appendix A, 
the steady state probability distribution can still be written in a factorized form
\bea
P(\{ n_{i_1} \}, \{ n_{i_2} \}) = \frac{1}{Z_N} \left[ \prod_{i_1=1}^{L_1} 
f_1(n_{i_1}) \prod_{i_2=1}^{L_2} f_2(n_{i_2}) \right] \nonumber \\
\times e^{\tilde{\mu}_1 {N_1}} e^{\tilde{\mu}_2 {N_2}} \delta(N_1+N_2-N)
\label{ZRP_steady_state2}
\eea
where ${\tilde{\mu}_1}= \ln({1}/{v_1})$ and ${\tilde{\mu}_2}= \ln({1}/{v_2}$), which we call excess chemical potentials. Now 
the joint probability distribution $P(N_1, N_2)$ of particle-numbers $N_1$ and $N_2$ can be expressed as 
\be
P(N_1, N_2) = \frac{Z_1(N_1) Z_2(N_2)}{Z_N} e^{\tilde{\mu}_1 {N_1}} e^{\tilde{\mu}_2 {N_2}}.
\ee 
The macrostate is obtained by maximizing $\ln P(N_1, N_2)$, i.e., $\partial \ln P(N_1, N_2)/\partial N_1=0$. Therefore it 
straightforwardly follows that
\be
\left( \frac{\partial \ln Z_1}{\partial N_1} + \tilde{\mu}_1 \right) = \left( \frac{\partial \ln Z_2}{\partial N_2} + 
\tilde{\mu}_2 \right)
\label{Excess_mu1}
\ee 
or, in other words, there indeed exists new intensive variable $\mu'_{\alpha}=(-\partial \ln Z_{\alpha}/\partial N_{\alpha} + 
\ln v_{\alpha})$, with $\alpha=1,2$, which takes the same values for two rings in contact, i.e., $\mu'_1=\mu'_2$. 
For the special case when $v_1=v_2$, the excess chemical potentials are equal (i.e., $\tilde{\mu}_1=\tilde{\mu}_2$) and 
drop out of Eq. \ref{Excess_mu1} which then implies that the old variables $\mu_1$ and $\mu_2$ equalize. But in the case 
when $v_1 \ne v_2$, 
the new variable $\mu_{\alpha}'$ takes the role of chemical potential which then equalize upon contact. 
This identification of $\mu_{\alpha}'$ as an intensive variable which equalizes for two systems in contact was missed in 
\cite{Bertin_PRL2006, Bertin_PRE2007} where it was concluded that there is no such equalization when $v_1 \ne v_2$.  
Moreover, in contrary to the suggestion in \cite{Bertin_PRE2007}, the detailed balance condition is still satisfied at the 
contact even when $v_1 \ne v_2$.

\begin{figure}[h!]
\begin{center}
\includegraphics [scale=0.65] {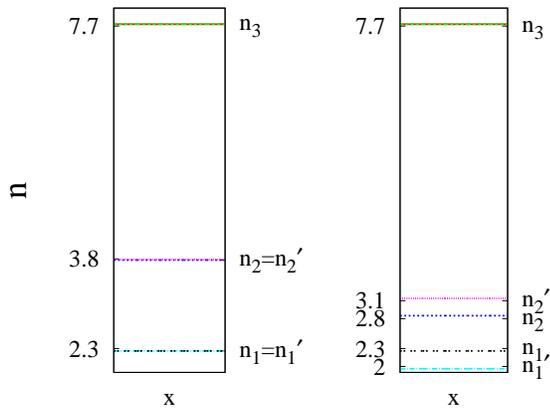}
\caption{ Density profiles for rings 1, 2 and 3 (all with same size $L=100$). Left panel: Three numerical experiments
to test the zeroth law in the case of the ZRP. Experiment 1: Ring 1 ($\delta_1=3$, $v_1=1$) in contact with ring 3 
($\delta_2=5$, $v_2=0.5$). Experiment 2: Ring 2 ($\delta_1=4$, $v_1=0.75$) in contact with ring 3 ($\delta_2=5$, 
$v_2=0.5$). Experiment 3: Ring 1 ($\delta_1=3$, $v_1=1$) in contact with ring 2 ($\delta_2=4$, $v_2=0.75$). 
The zeroth law is satisfied in this case.
Right panel: Three numerical experiments to illustrate the role of contact dynamics for the deviations from the zeroth law.
Experiment 1: Ring 1 ($\delta_1=3$, $v_1=1$) in contact with ring 3 ($\delta_2=5$, $v_2=0.5$). Experiment 2 with slightly
{\it perturbed jump rates} at the contact: Ring 2 ($\delta_1=4$, $v_1=0.85$) in contact with ring 3 ($\delta_2=5$, 
$v_2=0.4$). Experiment 3: Ring 1 ($\delta_1=3$, $v_1=1$) in contact with ring 2 ($\delta_2=4$, $v_2=0.75$).
The zeroth law is not satisfied in this case.
}
\label{0thLaw_vio_ZRP1}
\end{center}
\end{figure}

For a class of systems specified by a particular set of parameters $\{f_{\alpha}(n), v_{\alpha} \}$, it can be immediately
checked that the zeroth law is indeed satisfied.  In Fig. \ref{0thLaw_vio_ZRP1}, we perform various numerical experiments 
similar to those discussed
in Fig. \ref{zeroth_law_attractive} and \ref{zeroth_law_repulsive}. We choose $f_{\alpha}(n)=n^{\delta_{\alpha}-1}$ 
with various sets of parameter values for $\{ \delta_{\alpha}, v_{\alpha}\}$. 
In the first experiment, ring 1 with $\delta_1=3$, $v_1=1$ and density $n_1 \simeq 2.3$ equilibrated with ring 3 
with $\delta_2=5$, $v_2=0.5$ and density $n_3 \simeq 7.7$. In the second experiment, ring 2 with $\delta_1=4$, $v_1=0.75$ 
and density $n_2 \simeq 3.8$
is equilibrated with ring 3 with $\delta_2=5$, $v_2=0.5$ and density $n_3 \simeq 7.7$. In the third experiment, ring 1 
with $\delta_1=3$, $v_1=1$ and density $n_1' \simeq 2.3$ is equilibrated with ring 2 with $\delta_2=4$, $v_2=0.75$ and 
density $n_2' \simeq 3.8$ where $n_1' \simeq n_1$ and $n_2' \simeq n_2$. In this case one can see that the zeroth law is 
satisfied (see the left panel of Fig. \ref{0thLaw_vio_ZRP1}). We have also checked that values of the densities are such 
that chemical potential $\mu_{\alpha}'$ takes equal values for two rings in contact. In the right panel of Fig. 
\ref{0thLaw_vio_ZRP1}, we again perform three similar numerical experiments but now, in the second experiment, the values 
of the factors $v_1=0.85$ and $v_2=0.4$ are slightly perturbed by an arbitrary amount from the earlier values of the factors 
$v_1=0.75$ 
and $v_2=0.5$ which were chosen in the second experiment of the first set of numerical experiments (i.e., corresponding to 
the left panel of Fig. \ref{0thLaw_vio_ZRP1}). One can see that, in this case, the zeroth law is violated even though the 
factorization property as given in Eq. \ref{ZRP_steady_state2} exactly holds in each of the three experiments. 
This illustrates the role of the contact dynamics for the zeroth law which is satisfied only for a precise set of 
jump rates at the contact with no arbitrariness is allowed in these rates.

\subsection{B. Excess Chemical Potential in the Equilibrium KLS Model}

The above situation in the case of the zero range process (ZRP) is very similar to the special case of the equilibrium KLS 
model when one chooses the 
transition rates as given in Eqs. \ref{Metropolis1} and \ref{Metropolis2} with no driving in the bulk of 
the individual systems, i.e., $E_1=E_2=0$, and a slightly modified transition rates at the contact as follows. In general,
the transition rate for a particle jumping from system 1 to system 2 is given by 
\be
w(C \rightarrow C') = v_1 \times \mbox{min}\{1, e^{- \Delta(0)}\}
\ee
and the reverse transition rate for the particle jumping from system 2 to system 1 is given by 
\be
w(C' \rightarrow C) = v_2 \times \mbox{min}\{1, e^{-\Delta(0)}\}
\ee 
where we put $\beta=1$. For $v_1=v_2=1$, the jump rates at the contact are same as given in Eq. \ref{Metropolis2}. Clearly 
the modified transition rates at the contact amount to an additional field $E_{12}=\ln (v_1/v_2)$, say from system 1 to 
system 2, along the bonds connecting the two systems. Note that the field $E_{12}$ is not a driving field and just introduces 
an extra overall shift in the chemical potential of the system 1. The field $E_{12}$ accordingly modifies the energy 
function from $H$ to $H'$ by introducing an extra chemical potential in Eq. \ref{H1}, i.e., $H'= H + E_{12} N_1$. 
In this case, the detailed balance is satisfied with respect to the Boltzmann distribution with the modified energy 
function $H'$. Consequently, one could effectively think of an excess contribution to the equilibrium free energy of the system 
1 due to the shift in the chemical potential of the system 1 by an amount $E_{12}$. Now the condition of minimization of the 
total free energy $F=F_1+F_2$, as given in Eq. \ref{Excess_mu1} for the zero range process,
would imply that the new intensive variables $\mu'_{\alpha}=(-\partial \ln Z_{\alpha}/\partial N_{\alpha} + 
\ln v_{\alpha})$, not the variables $\mu_{\alpha}=-\partial \ln Z_{\alpha}/\partial N_{\alpha}$, equalize where 
free energies of the respective systems are $F_1=-\ln Z_1+\tilde{\mu}_1 N_1$ and $F_2=-\ln Z_2+\tilde{\mu}_2 N_2$ with partition 
function $Z_{\alpha}=\sum_C e^{-H_{\alpha}(C)}$, the energy function for 
individual system $H_{\alpha}=K_{\alpha} \sum_{\langle {\bf r_{\alpha}}, {\bf r_{\alpha}'} \rangle} \eta({\bf r_{\alpha}}) 
\eta({\bf r_{\alpha}'})$ (see Eq. \ref{H1}) for $\alpha=1,2$. Consequently, the zeroth law is satisfied in the case
when each system is assigned a particular set of values of the factors $v_{\alpha}$.
However, if the factors $v_{\alpha}$ are chosen arbitrarily in any of the numerical experiments as discussed in
the right panel of Fig. \ref{0thLaw_vio_ZRP1} in the case of the ZRP, the zeroth law would not hold even for the equilibrium 
KLS model with the modified jump rates at the contact.

\section{V. Analytical Approaches for the Driven KLS Model}

In this section, first we discuss how the zeroth law, which has been seen to be satisfied to a very good approximation in the 
simulations of the driven KLS model for various parameter values, could be explained in terms of large deviation principle. 
Then we discuss how the deviations from the zeroth law, also observed in the simulations, could be explained by modifying this 
large deviation principle.

\subsection{A. Large Deviation Principle}

In equilibrium, the zeroth law can be derived from variational principles, e.g., by maximization of entropy of an 
isolated system or by minimization of free energy in the case when the system is in contact with a reservoir. 
For some nonequilibrium systems, there may be a similar principle, called the large deviation principle 
\cite{Eyink1996, Touchette}. Then, the zeroth law can be derived from the large deviation principle along the same 
line as in equilibrium in the following way. Let us consider the following scenario. Two systems are kept in contact with 
each other with a particular dynamics specified at the contact. The two systems exchange according to the contact dynamics
some conserved quantity, say number of
particles, such that $N_1+N_2 = N = constant$ with $N_1$, $N_2$ being the number of particles 
in systems 1 and 2 respectively (schematically shown in Fig. \ref{dia_contact}). The quantities $N_1$, $N_2$ are 
considered to be extensive, i.e., proportional to the volume $V_1$, $V_2$ of system 1 and 2 respectively. Now we are 
interested in large deviations of $N_1$ and $N_2$ and assume that the probability of large deviation
$P(N_1, N_2)$ in the quantity $N_1$ and $N_2$ is given by 
\be 
P(N_1, N_2) \sim \frac{e^{-V_1 f_1(n_1)} e^{-V_2 f_2(n_2)}}{e^{-F(N)}}
\label{large_deviation_principle} 
\ee 
in the limit of $N_1, N_2, V_1, V_2 \gg 1$ keeping respective densities $n_1=N_1/V_1$ and $n_2=N_2/V_2$ finite with 
the normalization constant $\exp[-F(N)]$. The above equation is the statement of the large deviation 
principle \cite{Eyink1996, Touchette} and the functions $f_1(n_1)$, $f_2(n_2)$ are called the large deviation functions 
for the corresponding systems. The sign `$\sim$' implies equality in terms of logarithm of the
respective quantities and Eq.\ref{large_deviation_principle} can be written more rigorously as
\bea 
-\ln P(N_1, N_2) = V_1 f_1(n_1) + V_2 f_2(n_2) - F(N) \nonumber
\\ + \epsilon(N_1, N_2)
\label{LDF_rigorous}
\eea 
where $\epsilon(N_1, N_2)/\ln P(N_1, N_2) \rightarrow 0$ as $N_1, N_2 \rightarrow \infty$. Note that, in writing Eqs. 
\ref{large_deviation_principle} and \ref{LDF_rigorous}, we have assumed that the correlation between system 1 and 2 can be 
neglected as a boundary-effect in the limit of large volume. The macroscopic stationary state $\{N_1^*, N_2^*\}$, under the 
constraint $N_1 + N_2 = constant$, can be determined by maximizing $\ln P(N_1, N_2)$, i.e., 
$\partial \ln P(N_1, N_2)/\partial N_1 = 0$ which gives 
\be
\left( \frac{\partial f_1}{\partial n_1} \right)_{n_1^*} = \left(
\frac{\partial f_2}{\partial n_2} \right)_{n_2^*} = \mu
\label{def_intensive_variable} 
\ee 
where $n_1^*=N_1^*/V_1$,
$n_2^*= N_2^*/V_2$ and $\mu$ is the chemical potential which takes same values in the steady state when system 1 and 2 
are kept in contact. Note that the validity of Eq. \ref{def_intensive_variable}, which implies the existence of an 
intensive thermodynamic variable, follows from the assumption of a large deviation principle of 
Eq. \ref{large_deviation_principle}. The above arguments are quite general and could be valid irrespective of a specific 
nature of the dynamics one considers in a particular problem. Therefore Eq. \ref{def_intensive_variable} can be equally 
applicable to an equilibrium as well as a nonequilibrium steady state. Another consequence of equalization of intensive 
variables is the fluctuation-response relation between the compressibility and the fluctuations in particle-number $N_1$ 
of system 1, i.e.,
\be
\frac{\partial \langle N_1 \rangle}{\partial \mu} = \langle N_1^2 \rangle - \langle 
N_1 \rangle^2,
\ee
when system 1 is kept in contact with a very large system 2, with chemical potential $\mu$, which can be considered as 
a particle reservoir (i.e., $N_1 \ll N_2$ and $V_1 \ll V_2$). For more details regarding the above fluctuation relation
in the context of the ZRP as well as for the KLS model, see Appendix B.

\subsection{B. Modified Large Deviation Principle}

The exact results for the ZRP and the equilibrium KLS model give us an insight into the role of the contact dynamics which 
can effectively introduce an excess chemical potential across the contact region. Consequently, the intensive variable for
a system 
which equalizes upon contact can have different functional forms depending on the specifics of the dynamics at the contact 
and therefore on the other system in contact. To describe this situation in a more general context where there may be 
equalization of an intensive variable but still the zeroth law does not strictly hold, we write the large deviation 
probabilities as given in Eq. \ref{large_deviation_principle} in a modified asymptotic form
\be 
P(N_1, N_2) \sim \frac{e^{-V_1 f_1(n_1)} e^{-V_2 f_2(n_2)} 
e^{\tilde{\mu}_1 N_1} e^{\tilde{\mu}_2 N_2} }{e^{-F(N)}}
\label{large_deviation_principle2} 
\ee
where $f_1$, $f_2$ are the large deviation functions for a putative contact dynamics for which the zeroth law is satisfied, 
the two additional 
quantities $\tilde{\mu}_1$, $\tilde{\mu}_2$ can be thought of as excess chemical potentials arising solely due to the 
actual contact dynamics which is under consideration and for which the zeroth law is not satisfied. Clearly, for an 
arbitrarily chosen contact dynamics, the old intensive variables $\mu_1=\partial f_1/\partial n_1$ and $\mu_2 = 
\partial f_2/\partial n_2$ do not equalize. Note that, unlike in the case of the ZRP or the equilibrium KLS model, the 
potentials $\tilde{\mu}_1$ and $\tilde{\mu}_2$ is in general a-priori unknown for a given contact dynamics. 
Moreover, for an arbitrarily chosen contact dynamics, it may not always be possible to assign, to the individual systems, 
an intensive variable which is independent of the specifics at the contact between two systems. In this case, even though there 
can be equalization of some intensive variable when two systems are kept in contact, the zeroth law may not be satisfied as
demonstrated in the case of the ZRP (see right panel of Fig. \ref{0thLaw_vio_ZRP1}).

The above modified large deviation principle in Eq. \ref{large_deviation_principle2} still assumes that two systems in 
contact have an asymptotic factorization property in the sense that the 
correlations between the systems can be ignored in the large volume limit. In a special case (e.g., the ZRP or the 
equilibrium KLS model), the chemical potentials $\tilde{\mu}_1$ and $\tilde{\mu}_2$ can be constant over an entire density 
range. However, in general, the quantities $\tilde{\mu}_1 (n_1)$ and $\tilde{\mu}_2(n_2)$ can be functions of densities 
$n_1$ and $n_2$. Note that, although the large deviation principles in Eqs. \ref{large_deviation_principle} and 
\ref{large_deviation_principle2} appear to have different forms, they are essentially the same. The only difference is that, 
in the modified form of Eq. \ref{large_deviation_principle2}, we identify the contribution to the large deviation function due 
to the contact dynamics and separate this contribution from that arising due to the bulk of the individual systems.

\begin{figure}[h!]
\begin{center}
\includegraphics [scale=0.70] {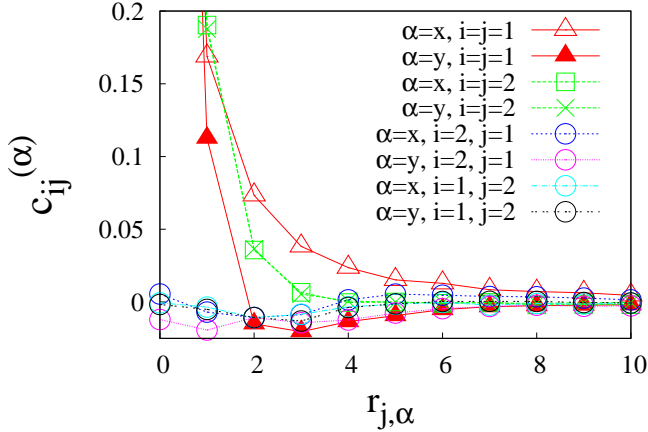}
\caption{Correlations between various neighboring sites as a function of distance when a driven system 1 
($K=-0.75$, $E=4$, $V=50 \times 50$) and an equilibrium system 2 ($K=-0.75$, $E=0$, $V=50 \times 50$) in contact 
with each other. The function $c^{(\alpha)}_{ij}(r_{j, \alpha})$ denotes the scaled correlation (see Eq. \ref{scaled_cor}) 
along $\alpha$th direction between two points located in system $i$ and system $j$ where $r_{j, \alpha}$ being the $\alpha$th 
component of relative position vector $\bf{r}_j$ between the two points with $i,j=1,2$ and $\alpha=x, y$ (for two 
dimensional system), $\Delta \eta ({\bf r}) = \eta({\bf r}) - \langle \eta({\bf r}) \rangle$ the fluctuation in occupation 
variable $\eta({\bf r})$. For $i \ne j$, $c^{(\alpha)}_{ij}(r_{j, \alpha}=0)$ is the correlation between two nearest-neighbor 
sites located across the contact.
}
\label{correlation1}
\end{center}
\end{figure}

The asymptotic factorization can indeed be a very good approximation even in the case of a more complicated driven
lattice gas like the KLS model. 
To show this, we now study various spatial density-correlations $c^{(\alpha)}_{ij}(r_{j, \alpha})$ between 
two points located in the individual systems as well as two points located in two different systems across the contact. 
We define the function 
\be
c^{(\alpha)}_{ij}(r_{j, \alpha}) = 
\frac{\langle \{ \Delta \eta ({\bf r_i^c}) \} \{\Delta \eta ({\bf r_j^c}+{\bf r_j})\}  \rangle}{
\sqrt{\langle\{\Delta \eta ({\bf r_i^c})\}^2\rangle} \sqrt{\langle\{\Delta \eta ({\bf r_j^c}+{\bf r_j})\}^2\rangle} },
\label{scaled_cor}
\ee
which denotes the scaled correlation along $\alpha$th direction between two points in system $i$ and system $j$ where 
$r_{j, \alpha}$ the $\alpha$th component of position vector $\bf{r}_j$ with $i,j=1,2$ and $\alpha=x, y$ (for a 
two-dimensional system), $\Delta \eta ({\bf r}) = \eta({\bf r}) - \langle \eta({\bf r}) \rangle$ the fluctuation in 
occupation variable $\eta({\bf r})$, ${\bf r_i^c}$ the position vector of the contact site in system $i$. For $i=j$, 
$c^{(\alpha)}_{ii}(r_{i, \alpha})$ denotes density-correlations, along $\alpha$-axis, between two points both located 
in the same system $i$ and, for $i\ne j$, $c^{(\alpha)}_{ij}(r_{j, \alpha})$ denotes cross-correlation, along $\alpha$-axis, 
between two points one located in the system $i$ 
and the other located in system $j$. In Fig. \ref{correlation1}, we have plotted the scaled correlation 
$c^{(\alpha)}_{ij}(r_{j, \alpha})$ as a function of $r_{j, \alpha}$ for a driven system with $K_1=-0.75$, $E_1=4$ and 
an equilibrium 
system with $K_2=-0.75$ where the systems are kept in contact, both at density $n=1/2$. One can see that the amplitude of the 
cross-correlations among nearest-neighbor sites located in two different systems across the contact are very small, almost an 
order of magnitude smaller as compared 
to those among neighboring sites in the individual systems, i.e., $c^{(\alpha)}_{21}(0), c^{(\alpha)}_{12}(0) \ll 
c^{(\alpha)}_{11}(1), c^{(\alpha)}_{22}(1)$. Therefore the asymptotic factorization property is expected to be well-satisfied
in this case. Moreover, the very weak cross-correlation between two systems in contact explains why the density profiles 
remain almost homogeneous even around the contact. Note that the density-correlation functions $c^{(x)}_{11}(r_{1,x})$ and 
$c^{(y)}_{11}(r_{1,y})$ for the driven system along $x$ and $y$ directions, respectively, are clearly different due to the 
presence of a strong driving field which breaks the isotropy. Whereas, for the equilibrium system, correlations 
$c^{(x)}_{22}(r_{1,x})$ and $c^{(y)}_{22}(r_{1,y})$ expectedly remain the same. For attractive interaction strengths, the 
correlations perpendicular to the driving field become negative at larger distances. This picture remains qualitatively the 
same at other densities as well.

\subsection{C. Excess Chemical Potential in the Driven KLS Model}

The role of the contact dynamics in the KLS model can be more complex than that in the previously discussed models of the ZRP. 
However, interestingly for some parameter values, we indeed find evidence of a constant difference 
$\delta \mu = \tilde{\mu}_1 - \tilde{\mu}_2$ in the excess chemical potentials in a wide density range 
even in the KLS model for nonzero driving. This supports the modified form of the large deviation principle even for the
KLS model.

\begin{figure}[h!]
\begin{center}
\includegraphics [scale=0.70] {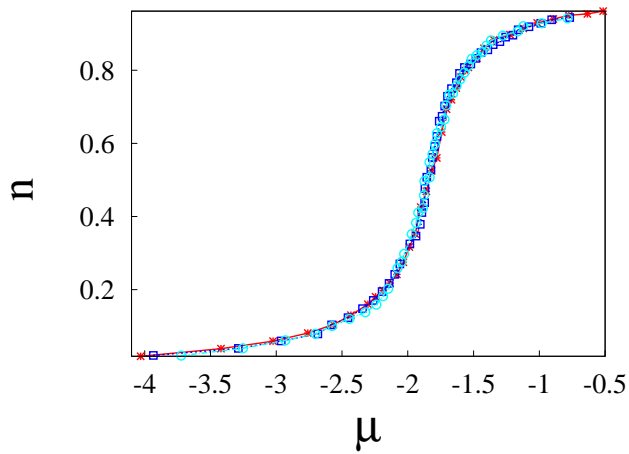}
\caption{Density $n$ {\it vs.} shifted chemical potential $\mu$ is plotted for
a driven system with $K=-1$, $E=6$ which is separately kept in contact with 
three equilibrium systems with interaction strengths $K=0$, $K=-0.8$, and $K=-0.9$. 
By shifting $\mu \rightarrow \mu + \delta \mu$, all curves collapse on each other well 
where we choose $\delta \mu = 0.06$ and $\delta \mu = 0.095$ for the cases when the driven system
is in contact with the equilibrium systems with $K=-0.8$ and $K=-0.9$, respectively.}
\label{excess_mu1}
\end{center}
\end{figure}

We consider density $n$ as a function of chemical potential $\mu$ for a driven system with $K=-1$, $E=6$ which is separately 
kept in contact with various equilibrium systems with interaction strengths $K=0$, $K=-0.8$, and $K=-0.9$. As we have already 
discussed in Fig. \ref{0thLaw_vio1} for the same systems, there were indeed deviations from the zeroth law observed as 
the different curves for $n$ {\it vs.} $\mu$ in Fig. \ref{0thLaw_vio1} do not fall on each other. However, now by shifting 
chemical potential $\mu$ to $\mu + \delta \mu$ and choosing a suitable $\delta \mu$ in each of the cases, all the 
curves can be made to collapse on each other quite well as can be seen in Fig. \ref{excess_mu1}.

\begin{figure}[h!]
\begin{center}
\includegraphics [scale=0.70] {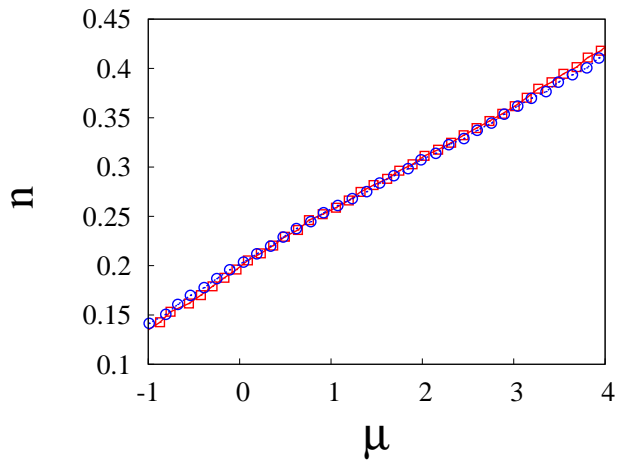}
\caption{Density $n$ {\it vs.} shifted chemical potential $\mu$ is plotted for
a driven system with $K=3.75$, $E=6$ which is separately kept in contact with 
two equilibrium systems with $K=0.75$ and $K=1$. 
Shifting $\mu \rightarrow \mu + \delta \mu$, two curves collapse on each other 
reasonably well by choosing $\delta \mu = 0.1$ for the case when the driven system is in 
contact with the equilibrium system with $K=0.75$.}
\label{excess_mu2}
\end{center}
\end{figure}

We have done the same analysis for another driven system with a different set of parameter values $K=3.75$, $E=6$. The 
driven system is separately kept in contact with two equilibrium systems with interaction strengths $K=0.75$ and $K=1.0$. In 
Fig. \ref{excess_mu2}, the densities $n$ are plotted as a function of the shifted chemical potential $\mu$ (also see Fig. 
\ref{0thLaw_vio2}). The curves could be collapsed on top of each other reasonably well by shifting the 
$\mu$ to $\mu + \delta \mu$.

However, in both the cases of Figs. \ref{excess_mu1} and \ref{excess_mu2}, it should be noted that the collapse is not so 
good at very low chemical potentials (i.e., low densities). This indicates that the difference in the excess chemical potential 
$\delta \mu$ actually may not be  
constant over the entire chemical potential range and can depend on the chemical potential itself (or equivalently density) of 
the corresponding equilibrium system in contact.

\section{VI. Concluding Perspective}

In summary, in this paper we have studied ``equilibration'' of driven lattice gases when two systems are kept in 
contact and allowed to exchange particles with the total number of particles conserved. Interestingly, both for 
attractive as well as for repulsive nearest-neighbor interactions and for a wide range of parameter values, there is a 
remarkably simple, though approximate, thermodynamic structure where the zeroth law is quite well satisfied.

However, there are also observed deviations from this simple thermodynamic law. To understand these deviations, first we have 
studied the nontrivial role of the contact dynamics for a variant of the zero range process (ZRP) as well as a variant of the 
equilibrium Katz-Lebowitz-Spohn (KLS) model, where one can calculate the steady-state probability distribution exactly. Using 
these simple examples, we point out that, due to the modified contact dynamics, there can be an excess chemical potential 
induced across the contact region. These results lead us to express the large 
deviation principle in a modified form which elucidates the role of the contact and which can substantially 
account for the deviations from the zeroth law. 

It is important to note that the modified form of the 
large deviation principle is still based on the asymptotically factorized form of the steady-state distribution for two driven 
systems in contact and is valid if the correlations between the systems can be neglected in the large volume limit.
In the case of the variant of the ZRP discussed here, these correlations are zero. 
In the case of the driven KLS model, we observed that the spatial density-correlations across the contact are indeed very
small compared to the nearest-neighbor correlations in the individual systems. Interestingly, we found evidence of 
an almost constant excess chemical potential for various parameter values and in a wide range of densities, therefore supporting 
applicability of the modified large deviation principle to the KLS model.

In general, the results in this paper lead us roughly to the following possible scenarios for driven systems in contact.
\\ 
(1) The large deviation principle, or in other words the asymptotic factorization property, may break down due to long-ranged 
correlations which may be present in the driven systems \cite{Dorfman, Garrido, Grinstein}. In this case, the combined system 
cannot be divided into independent subsystems and there would be no intensive variable which equalizes upon contact. 
\\
(2) However, when the amplitude of these correlations are sufficiently weak, it is possible that a large deviation principle 
holds, although in a modified form. In these cases, the systems can be characterized by the excess chemical potentials across 
the contact and consequently there would be some intensive variable which would then equalize upon contact. Note that 
introducing the excess chemical 
potential is essentially a way of reparametrization of the chemical potential of the driven system under consideration. These 
excess chemical potentials can depend on the specifics of the contact dynamics and they are generally a-priori unknown, therefore 
in a sense arbitrary. For some parameter values, the excess chemical potential may almost be constant over a range of densities. 
In special cases, the arbitrariness of the excess chemical potential can be removed by choosing a suitable contact dynamics 
such that the zeroth law holds strictly. In these cases, it is actually possible to assign to the individual systems an intensive 
variable, independent of the specifics at the contact between two systems, which would then equalize.
It should be noted here that the modified large deviation principle could be satisfied irrespective of whether the zeroth law 
is satisfied or not, which has been illustrated in this paper by using a variant of the ZRP.

For the ZRP, since the steady state properties are exactly known, it is easier to choose a contact dynamics so that the zeroth 
law can be made to be satisfied. However, for the KLS model which has nontrivial steady state properties, it 
would be difficult to find a contact dynamics, even if it exists, for which the zeroth law would hold strictly.
For the KLS model, here we have considered a contact dynamics, mainly based on a physical ground 
albeit on an ad-hoc basis, which satisfies the local detailed balance condition. The simple modification of the large deviation 
principle suggested in this paper indicates that it could still be possible to choose a contact dynamics such that the 
thermodynamic laws are satisfied even better.  
As an open question, the origin of the excess chemical potential at the contact should be understood in more detail which could 
give valuable insights whether it is possible to choose a contact dynamics for which a simple thermodynamic structure emerges for 
driven systems in general.

There are further important aspects in exploring such a simple structure for driven systems in contact. Unlike in equilibrium 
where there is a well defined prescription to describe various thermodynamic properties using the standard Boltzmann distribution, there is no such prescription for nonequilibrium systems. However, the numerically observed simple thermodynamic structure 
concerning driven lattice gases in contact may give us an useful tool to characterize such systems and gives rise to a 
possibility to describe phase transitions which are known to occur in these driven interacting many-particle systems.

\section{Acknowledgments}

We thank R. K. P. Zia for stimulating discussions.

\section{Appendix A: Steady state distribution for the ZRP when $v_1 \ne v_2$}

In the general case  when $v_1 \equiv v^{(c)}_1 \ne v_2 \equiv v^{(c)}_2$, the ansatz for the steady-state probability 
distribution is given by
\bea
P_{st}(\{ n_{i_1} \}, \{ n_{i_2} \}) = \frac{1}{Z_N} \left[ \prod_{i_1=1}^{L_1} 
f_1(n_{i_1}) \prod_{i_2=1}^{L_2} f_2(n_{i_2}) \right] \nonumber \\
\times \frac{1}{v_1^{N_1} v_2^{N_2}}  \delta(N_1+N_2-N) \mbox{~~~~}
\label{ZRP_steady_state3}
\eea
where $N_1$ and $N_2$ are number of particles in ring 1 and ring 2 respectively and $N=N_1+N_2$
is the total conserved particle number. Now we consider the following two cases - (1) when particle jumps 
in the bulk, say in ring 1, and (2) when particle jumps from one ring to the other.
\\
\\
Case (1): Consider the following two transitions from a configuration $C$ to $C'$ and 
a configuration $C''$ to $C$ where
$C \equiv (\{\dots, n_{i_1-1}, n_{i_1}, n_{i_1+1}, \dots \}, \{n_{i_2}\})$, $C' \equiv (\{\dots, 
n_{i_1-1}, n_{i_1}-1, n_{i_1+1}+1, \dots \}, \{n_{i_2}\})$ and $C'' \equiv (\{\dots, n_{i_1-1}+1, 
n_{i_1}-1, n_{i_1+1}, \dots \}, \{n_{i_2}\})$. 
The steady-state probability current $J(C \rightarrow C') = P_{st}(C) w(C'|C)$ from 
configuration $C$ to $C'$, $P_{st}(C)$ being the probability of configuration $C$ in steady state 
and $w(C'|C)$ being the transition rate from $C$ to $C'$, can be explicitly written as
\bea
J(C \rightarrow C') = \frac{1}{Z_N} \left[ \dots f(n_{i_{1}-1}) f(n_{i_1}) f(n_{i_1+1}) \dots \right] \nonumber \\
\times \left[ v^{(b)}_1 \frac{f(n_{i_1}-1)}{f(n_{i_1})} \right]  \left[ \prod_{i_2=1}^{L_2} f_2(n_{i_2}) \right] \nonumber \\
\times \frac{1}{v_1^{N_1} v_2^{N_2}} \delta(N_1+N_2-N) \mbox{~~}
\eea
where we have used jump rate in the bulk of ring 1 which is $w(C'|C) = v^{(b)}_1 {f(n_{i_1}-1)}/{f(n_{i_1})}$. 
Finally we get
\bea
J(C \rightarrow C') = \frac{1}{Z_N} \left[ \dots f(n_{i_{1}-1}) f(n_{i_1}-1) f(n_{i_1+1}) \dots \right] \nonumber \\
\times  \left[ \prod_{i_2=1}^{L_2} f_2(n_{i_2}) \right] \frac{v^{(b)}_1}{v_1^{N_1} v_2^{N_2}} \delta(N_1+N_2-N). \mbox{~~~~~}
\eea
Similarly the probability current $J(C'' \rightarrow C) = P(C'') w(C|C'')$ from 
configuration $C''$ to $C$ can be explicitly written as
\bea
J(C'' \rightarrow C) = \frac{1}{Z_N} \left[ \dots f(n_{i_{1}-1}+1) f(n_{i_1}-1) f(n_{i_1+1}) \dots \right] \nonumber \\
\times \left[ v^{(b)}_1 \frac{f(n_{i_1-1})}{f(n_{i_1-1}+1)} \right] \left[ \prod_{i_2=1}^{L_2} f_2(n_{i_2}) \right] \nonumber \\
\times \frac{1}{v_1^{N_1} v_2^{N_2}} \delta(N_1+N_2-N) \nonumber \\
= \frac{1}{Z_N} \left[ \dots f(n_{i_{1}-1}) f(n_{i_1}-1) f(n_{i_1+1}) \dots \right] \nonumber \\
\times  \left[ \prod_{i_2=1}^{L_2} f_2(n_{i_2}) \right] \frac{v^{(b)}_1}{v_1^{N_1} v_2^{N_2}} \delta(N_1+N_2-N). \mbox{~~~~~~~} 
\eea
Clearly the probability currents $J(C \rightarrow C')$ and $J(C'' \rightarrow C)$ are equal. Since, for any transition from $C$ to $C'$, it is always possible to find a corresponding transition $C''$ to $C$, the net current into $C$ vanishes pairwise, i.e.,
$J(C \rightarrow C')=J(C'' \rightarrow C)$ in the steady state. 
\\
\\
Case (2):  Consider two transitions from $C$ to $C'$ and the configuration $C'$ to $C$ where
$C \equiv (\{\dots, n_{k_1}, \dots \}, \{ \dots, n_{k_2}, \dots \})$, 
$C' \equiv (\{\dots, n_{k_1}-1, \dots \}, \{ \dots, n_{k_2}+1, \dots \})$, and 
$k_1$ and $k_2$ are respective contact sites in ring 1 and ring 2. In this case the probability 
current $J(C \rightarrow C')$ can be written as
\bea
J(C \rightarrow C') = \frac{1}{Z_N} \left[ \dots f(n_{k_1}) \dots \right] \left[ v_1 \frac{f(n_{k_1}-1)}{f(n_{k_1})} \right] \nonumber \\ 
\times \left[ \dots f(n_{k_2}) \dots \right] \frac{1}{v_1^{N_1} v_2^{N_2}} \delta(N_1+N_2-N) \nonumber \\
= \frac{1}{Z_N} \left[ \dots f(n_{k_1}-1) \dots \right] \left[ \dots f(n_{k_2}) \dots \right] \nonumber \\
\times \frac{1}{v_1^{N_1-1} v_2^{N_2}} \delta(N_1+N_2-N) 
\eea
Similarly the probability current $J(C' \rightarrow C)$ can be written as
\bea 
J(C' \rightarrow C) = \frac{1}{Z_N} \left[ \dots f(n_{k_1}-1) \dots \right] \left[ \dots f(n_{k_2}+1) \dots \right] \nonumber \\
\times  \left[ v_2 \frac{f(n_{k_2})}{f(n_{k_2}+1)} \right] \frac{1}{v_1^{N_1-1} v_2^{N_2+1}} \delta(N_1+N_2-N) \nonumber \\
=  \frac{1}{Z_N} \left[ \dots f(n_{k_1}-1) \dots \right] \left[ \dots f(n_{k_2}) \dots \right] \nonumber \\
\times \frac{1}{v_1^{N_1-1} v_2^{N_2}} \delta(N_1+N_2-N). \mbox{~~~~~~~} 
\eea
Clearly the net probability current into $C$ is again zero as $J(C \rightarrow C') = J(C' \rightarrow C)$ in the steady 
state. This completes the proof for the steady-state ansatz given in Eq. \ref{ZRP_steady_state3} which satisfies 
Master equation $\partial_t P_{st}(C,t) = 0 = \sum_{C' \ne C} [P_{st}(C') w(C' \rightarrow C)-P_{st}(C) w(C \rightarrow C')]$.
Note that the condition $J(C \rightarrow C') = J(C' \rightarrow C)$ is nothing but the detailed balance condition which is 
satisfied at the contact even when $v_1 \ne v_2$. However, detailed balance is not satisfied in the bulk and consequently
there are nonzero currents within the individual systems.

\section{Appendix B: Fluctuation-response Relations}

One interesting consequence of Eq. \ref{def_intensive_variable} is a relation between the susceptibility and the
fluctuation in particle-number of a system when the system is in contact with a large reservoir characterized by a chemical 
potential $\mu$. Consider system 1 is in contact with system 2 which is very large compared to system 1. Let us denote 
$\sigma_{N_1}^2$ as the standard deviation of fluctuations in the total number of particles $N_1$ in system 1, i.e., 
$\sigma_{N_1}^2 = \langle N_1^2 \rangle - \langle N_1 \rangle^2$. Then the large deviation principle with the definition 
of chemical potential as given in Eq. \ref{def_intensive_variable} implies that the fluctuation of particle-number around an 
average particle-number $N_1^*$, 
\be 
P(N_1) \approx \mbox{constant} \times e^{-(N_1-N_1^*)^2/2 \chi} 
\ee 
where $1/\chi = \partial \mu/\partial N_1 = (\partial^2 F_1/\partial N_1^2)_{N_1^*}$ with $F_1(N_1)=V_1 f_1(n_1)$. Since 
the root mean square fluctuation in the particle-number $N_1$ is $\sigma_{N_1} ^2 = 2 \chi$, one gets the following 
fluctuation relation, 
\be 
\frac{\partial \langle N_1 \rangle}{\partial \mu} =
(\langle N_1^2 \rangle - \langle N_1 \rangle^2) 
\label{FR1}. 
\ee

\begin{figure}[h!]
\begin{center}
\includegraphics [scale=0.70] {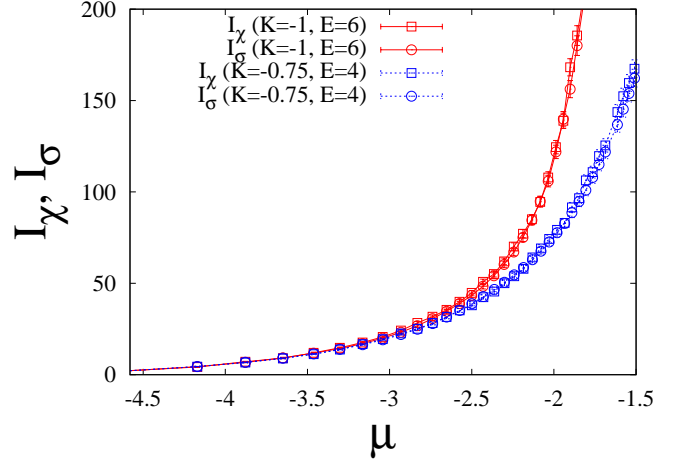}
\caption{Driven system in contact with an equilibrium reservoir of noninteracting hardcore particles.
Integrated compressibility $I_{\chi}$ (squares) and integrated fluctuation 
$I_{\sigma}$ (circles) plotted as a function of $\mu$ for a driven system with two different sets of parameter values
with $K=-1$, $E=6$ (red) and $K=-0.75$, $E=4$ (blue). The fluctuation-response relation is well satisfied. }
\label{FR_numerical1}
\end{center}
\end{figure}

For the ZRP, even when $v_1 \ne v_2$, the fluctuation-response relation between the compressibility and the root mean square 
fluctuation in particle-number $N_1$ is still exactly satisfied if the ring 1 is in contact with a much larger ring 2 being 
a particle reservoir. Moreover, since in this case the variables $\mu_{\alpha}'$ and $\mu_{\alpha}$ differ only by a constant 
$\ln v_{\alpha}$ (see Eq. \ref{Excess_mu1} and the paragraph below), the fluctuation-response relation is satisfied with 
respect to both the new and old intensive variables $\mu_{\alpha}'$ and $\mu_{\alpha}$, respectively.

Now we briefly discuss the numerical results concerning the fluctuation-response relation for the KLS model with 
attractive interaction strengths. To numerically test the fluctuation relation as given in Eq. \ref{FR1}, we consider a 
driven system which is in contact with an equilibrium reservoir of hardcore particles which are otherwise 
noninteracting (i.e., $K=0$). The chemical potential $\mu$ of the equilibrium hardcore particle-reservoir is given by the 
expression in Eq. \ref{mu}. 
For better numerical accuracy, we check the integrated version of Eq. \ref{FR1}, i.e., we calculate the integrated fluctuation
$I_{\sigma} = \int^{\mu}_{\mu_0} (\sigma_{N_1}^2) d \mu'$  and the integrated susceptibility $I_{\chi} = \int^{\mu}_{\mu_0} 
({d\langle N_1 \rangle}/{d\mu'}) d\mu'$ for different values of $\mu$ obtained by varying density of the equilibrium 
reservoir. We take a two-dimensional $20 \times 20$ nonequilibrium system in contact with a  $250 \times 250$ equilibrium 
reservoir. In Fig. \ref{FR_numerical1}, the integrated 
compressibility and the integrated fluctuations are plotted as a function of chemical potential $\mu$ for two driven 
systems with two different sets of parameter values with (1) $K=-1$, $E=6$ and (2) $K=-0.75$, $E=4$. In these cases,
the fluctuation-response relation is remarkably well satisfied as seen in Fig. \ref{FR_numerical1}.


\begin{thebibliography}{99}

\bibitem{Zia_NESS} R. K. P. Zia and B. Schmittmann, J. Stat. Mech. {\bf P07012} (2007).

\bibitem{Eyink1996} G. L. Eyink, J. L. Lebowitz, and H. Spohn, J. Stat. Phys. {\bf 83}, 385 (1996).

\bibitem{Oono_Paniconi1998} Y. Oono and M. Paniconi, Prog. Theor. Phys. Suppl. {\bf 130}, 29 (1998).

\bibitem{Bertini_etal} L. Bertini, A. D. Sole, D. Gabrielli, G. Jona-Lasinio, and C. Landim, 
Phys. Rev. Lett. {\bf 87}, 040601 (2001). 
L. Bertini, A. D. Sole, D. Gabrielli, G. Jona-Lasinio, and C. Landim, J. Stat. Phys. {\bf 107}, 635 (2002).

\bibitem{Bodineau_Derrida} T. Bodineau and B. Derrida, Phys. Rev. Lett. {\bf 92}, 180601 (2004).

\bibitem{Sasa2006} S. Sasa and H. Tasaki, J. Stat. Phys. {\bf 125}, 125 (2006).

\bibitem{Wang_Menon} H-Q Wang and N. Menon, Phys. Rev. Lett. {\bf 100}, 158001 (2008).

\bibitem{Henkes} S. Henkes, C. S. O'Hern, and B. Chakraborty, Phys. Rev. Lett. {\bf 99}, 038002 (2007).

\bibitem{Shokef} Y. Shokef, G. Shulkind, and D. Levine, Phys. Rev. E {\bf 76}, 030101(R) (2007).

\bibitem{Hayashi_Sasa2003} K. Hayashi and S. Sasa, Phys. Rev. E {\bf 68}, 035104 (2003).

\bibitem{Bertin_PRL2006} E. Bertin, O. Dauchot, and M. Droz, Phys. Rev. Lett. {\bf 96}, 120601 (2006). 

\bibitem{Evans} M. R. Evans and T. Hanney, J. Phys. A {\bf 38}, R195 (2005).

\bibitem{Majumdar} S. N. Majumdar, M. R. Evans, and R. K. P. Zia, Phys. Rev. Lett. {\bf 94}, 180601 (2005). 
M. R. Evans, T. Hanney, and S. N. Majumdar, Phys. Rev. Lett. {\bf 97}, 010602 (2006).

\bibitem{Bertin_PRE2007} E. Bertin, K. Martens, O. Dauchot, and M. Droz, Phys Rev. E {\bf 75}, 031120 (2007).

\bibitem{KLS} S. Katz et al., J. Stat. Phys. {\bf 34}, 497 (1984). S. Katz, J. L. Lebowitz, and H. Spohn, 
J. Stat. Phys. {\bf 34}, 497 (1984).

\bibitem{Zia} B. Schmittmann and R. K. P. Zia, Phys. Rep. {\bf 301}, 45 (1998).
R. K. P. Zia, J. Stat. Phys. {\bf 138}, 20 (2010).

\bibitem{Dietrich} W. Dietrich, P. Fulde, and I. Peschel, Adv. Phys. {\bf 29}, 527 (1980).

\bibitem{review1} J. Marro, J. L. Lebowitz, H. Spohn, and M. H. Kalos, J. Stat. Phys. {\bf 38}, 725 (1985).
J. L. Valles and J. Marro, J. Stat. Phys. {\bf 43}, 441 (1986). {\it ibid} {\bf 49}, 89 (1987).
K.-t Leung, Phys. Rev. Lett. {\bf 66}, 453 (1991). J-S Wang,  J. Stat. Phys. {\bf 82}, 1409 (1996).

\bibitem{Leung-Schmittmann-Zia} K.-t Leung, B. Schmittmann, and R. K. P. Zia, Phys. Rev. Lett. {\bf 62}, 1772 (1989).

\bibitem{review2} R. Dickman, Phys. Rev. A {\bf 38}, 2588 (1988). {\it ibid} {\bf 41}, 2192 (1990).

\bibitem{review3} H. K. Janssen and B. Schmittmann, Z. Phys. B - Cond. Matt. {\bf 64}, 503 (1986). 
K.-t Leung and J. L. Cardy, J. Stat. Phys. {\bf 44}, 567 (1986).

\bibitem{Pradhan_PRL} P. Pradhan, C. P. Amann, and U. Seifert, Phys. Rev. Lett. {\bf 105}, 150601 (2010).

\bibitem{Bilayr_systems} A. Achahbar and J. Marro, J. Stat. Phys. {\bf 78},
1493 (1995). C. C. Hill, R. K. P. Zia, and B. Schmittmann, Phys. Rev. Lett.
{\bf 77}, 514 (1996).

\bibitem{Touchette} H. Touchette, Phys. Rep. {\bf 478}, 1 (2009).

\bibitem{Dorfman} J. R. Dorfman, T. R. Kirkpatrick, and J. V. Sengers, Annu. Rev. Chem. {\bf 45}, 213 (1994).

\bibitem{Garrido} P. L. Garrido et al., Phys. Rev. A {\bf 42}, 1954 (1990).
P. L. Garrido, J. L. Lebowitz, C. Maes, and H. Spohn, Phys. Rev. A {\bf 42}, 1954 (1990).

\bibitem{Grinstein} G. Grinstein, D. H. Lee, and S. Sachdev, Phys. Rev. Lett. {\bf 64}, 1927 (1990).





\end{thebibliography}
\end{document}